\documentclass[12pt, breaklinks=true]{article}
\usepackage[top=1.5in, bottom=1.5in, left=1in, right=1in]{geometry}

\usepackage{enumerate}
\usepackage{ragged2e}
\usepackage{amssymb,array, amsmath,amsfonts, 
xcolor,graphicx, geometry, float}
\usepackage[T1]{fontenc}
\usepackage{libertine}
\usepackage{setspace}
\definecolor{darkred}{rgb}{0.65, 0, 0} 
\usepackage[
    colorlinks=true,        
    linkcolor=darkred,      
    citecolor=darkred,      
    urlcolor=darkred        
]{hyperref}

\allowdisplaybreaks[4]
\NewDocumentCommand{\abs}{o m}{
  \left\lvert
  \IfNoValueTF{#1}
    {#2}
    {\vphantom{#1}\smash{#2}}
  \right\rvert
}
\usepackage{threeparttable, tabularx}
\newcommand{\mc}[1]{\multicolumn{1}{c@{}}{#1}}
\newcolumntype{Y}{>{\centering\arraybackslash}X}
\usepackage{dcolumn} \newcolumntype{d}[1]{D..{#1}}\newcolumntype{L}{D{.}{.}{2,5}}
\usepackage{caption}
\usepackage[
    authordate,          
    backend=biber,       
    maxcitenames=3,      
    mincitenames=1,      
    sortcites=true,      
    annotation=true,     
    natbib=true,         
    doi=false,           
    url=false,           
    isbn=false,
    ibidtracker=false 
]{biblatex-chicago}
\DeclareFieldFormat{cmsnamehyper}{#1}
\DeclareCiteCommand{\citeyear}
  {\usebibmacro{cite:init}%
   \usebibmacro{prenote}}
  {\usebibmacro{citeindex}%
   \usebibmacro{citeyear}}
  {}
  {\usebibmacro{postnote}}
\DeclareCiteCommand{\citeyearpar}[\mkbibparens]
  {\usebibmacro{cite:init}%
   \usebibmacro{prenote}}
  {\usebibmacro{citeindex}%
   \usebibmacro{citeyear}}
  {}
  {\usebibmacro{postnote}}
\usepackage{appendix} 
\usepackage{booktabs,siunitx}
\usepackage{bookmark} 
\usepackage[labelfont=bf]{caption} 
\usepackage{indentfirst} 
\usepackage{pdflscape} 
\usepackage{siunitx} 
\usepackage{chngcntr} 
\usepackage{subfigure}
\usepackage{threeparttable}
\floatstyle{plaintop}
\restylefloat{table}
\usepackage{array, makecell}
\usepackage{multicol}
\usepackage{multirow}
\usepackage{lscape}
\usepackage{placeins}
\usepackage{csquotes}
\usepackage{setspace}
\graphicspath{{../fig/}{../figs/}{JEL_HF Paper_graphics/}{JEL_HF Paper_tcache/}{JEL_HF Paper_gcache/}}
\DeclareGraphicsExtensions{.pdf,.eps,.ps,.png,.jpg,.jpeg} 
\providecommand{\U}[1]{\protect \rule{.1in}{.1in}}

\input{tcilatex}
\usepackage{blindtext}

\usepackage{setspace}
\usepackage[none]{hyphenat}
\begin{document}
\sloppy

\title{\Large Homeownership as a Life-Cycle Goldmine: \\ Evidence from Macrohistory\thanks{We thank Ian Cherry (discussant), Aizhan Anarkulova (discussant), Shan Ge (discussant), and seminar and conference participants at the 2025 FMA Asia/Pacific Conference, 2025 MFA Annual Meeting, California State University, Fullerton, and the 2024 Boca Finance and Real Estate Conference for comments and suggestions. All errors are our own.}
}

{
\author{\large
Yang Bai\thanks{College of Business and Economics, California State University, Fullerton, \href{mailto:yabai@fullerton.edu}{yabai@fullerton.edu}.}\quad
Shize Li\thanks{Lingnan College, Sun Yat-sen University, \href{mailto:lishz5@mail.sysu.edu.cn}{lishz5@mail.sysu.edu.cn}.}\quad
Jialu Shen\thanks{International School of Finance, Fudan University, \href{mailto:jialus@fudan.edu.cn}{jialus@fudan.edu.cn}.} 
}
\date{ 
\small This draft: September 5, 2026 \\
First draft: July 31, 2024
}
}

\hypersetup{pageanchor=false}
\clearpage\maketitle
\thispagestyle{empty} 
\begin{center}
\vspace{-60pt}
\vspace{6pt}
\end{center}
\begin{abstract}
\singlespacing{
\noindent Should households buy their homes? Contrary to popular personal-finance and academic expert  advice, our block-bootstrap life-cycle simulation suggests they should. Using 150 years of data across 16 countries, we find that homeownership raises wealth and consumption-equivalent welfare relative to saving-rate-matched benchmark strategies investing solely in financial assets. The gains come from lower portfolio risk, rent-risk hedging, access to home equity late in life for enhanced retirement consumption, and higher bequests. The gains vary with income, house-price conditions at purchase, and mortgage rates. Mortgages enable early homeownership access at the cost of lower liquidity and consumption-equivalent welfare.

\vspace{24pt}

}
\end{abstract}
\vfill
\pagebreak
\setcounter{page}{1} 
\hypersetup{pageanchor=true}


\phantomsection
\pdfbookmark[1]{Introduction}{sec:introduction}
\noindent Homeownership is central to American households' portfolio allocation. Approximately 66\% of American households are homeowners \citep{census2024homeownership}, and 80\% of households entering homeownership finance the purchase \citep{nar2024generational}. Since 2014,  the public has perceived real estate   as the best long-run investment, beating stocks, bonds, and savings \citep{gallup2024investment}. As the most important asset class, housing is the primary means of storing wealth, especially for the bottom 50\% of the American wealth distribution \citep{diwan2021effects,campbell2006household}. The typical  homeowner's  home accounts for close to 80\% of  household assets \citep{campbell2025fixed}. 

Despite its popularity, housing as an investment instrument has long been criticized by experts. \citet{choi2022popular} documents that 14  of the 50 most popular personal finance books  state that housing is a mediocre investment, and \citet{cocharne2022portfolios} describes such investments  as ``disastrous.'' Additionally, houses' illiquid nature   could reduce household consumption and equity market participation, reducing life-cycle welfare \citep{campbell2006household}. 
In light of these conflicting views, it is crucial to study whether purchasing a home creates wealth and enhances household welfare in the  long run. We evaluate homeownership against common portfolio strategies, assuming the same saving rate under each strategy, and show that homeownership outperforms renting and investing solely in financial assets.

We apply the stationary block-bootstrap simulation \citep{politis1994stantionary} to the Jord\`{a}-Schularick-Taylor Macrohistory Database \citep{jorda2019rate} using a life-cycle model that incorporates labor income risk, risky returns from different asset classes, and longevity risk, through which we create a one-million-household panel and compare the performance of different life-cycle portfolio strategies under economic conditions that are identical for a given household but differ across households. 
Specifically, we consider a model in which  household members face separate labor-income and longevity risks. Each period during their working life, households earn labor income, pay for housing, deposit discretionary savings, and consume the rest of their income. When they retire, their income is replaced with   withdrawals from their financial portfolio to supplement their Social Security retirement benefits (SSRBs). Before retirement, homeowners who exhaust their financial wealth and fall below the minimum consumption level must liquidate their home to meet their consumption needs. Retired homeowners can also take out reverse mortgages. Renters in a similar situation receive   safety net income in the form of Supplemental Security Income (SSI) to make up the gap and remain at the minimum consumption level.

To evaluate homeownership's investment value, we simulate households repeatedly along their own lifetime economic paths but under different portfolio rules. In one life, a household follows a homeowner strategy with access to housing and liquid financial assets: They purchase a home when a down payment criterion and an additional purchase threshold are satisfied. In parallel lives, the household follows renter benchmarks that exclude housing and invest solely in financial assets, with the same discretionary saving rate as the homeowner's. Given a pair of homeowner and renter benchmark strategies, we average the difference in the household performance between them to assess the relative value of homeownership access across the one-million-household panel. 

We examine saving rates of 10\%, 15\%, and 20\%, following \citeauthor{dynan2004rich}'s  \citeyearpar{dynan2004rich} comprehensive household saving evidence, \citeauthor{huggett2000saving}'s \citeyearpar{huggett2000saving} empirical and model-implied optimal rate, and the range of popular advice surveyed by \citet{choi2022popular}. For brevity, we benchmark our main results against the all-equity renter strategy, which prior work identifies as the strongest life-cycle allocation \citep{anarkulova2023beyond,choi2025practical}. Relative to this benchmark, the leveraged homeowner strategy, with a 10\% down payment and a 10\% purchase threshold, achieves a 12.48\% terminal-wealth gain and a 3.23\% average household-level consumption-equivalent variation (CEV) at the baseline saving rate of 15\%.\footnote{We refer to a positive (negative) CEV as a consumption-equivalent gain (loss).} At this saving rate, cash-purchase strategies deliver the largest lifetime consumption-equivalent gains, reaching as high as 4.5\%. Since we do not impose a mechanical housing preference, our simulation reveals the lower bound of the consumption-equivalent gains from homeownership. 

\begin{figure}[t]
    \centering
    \includegraphics[width=\linewidth]{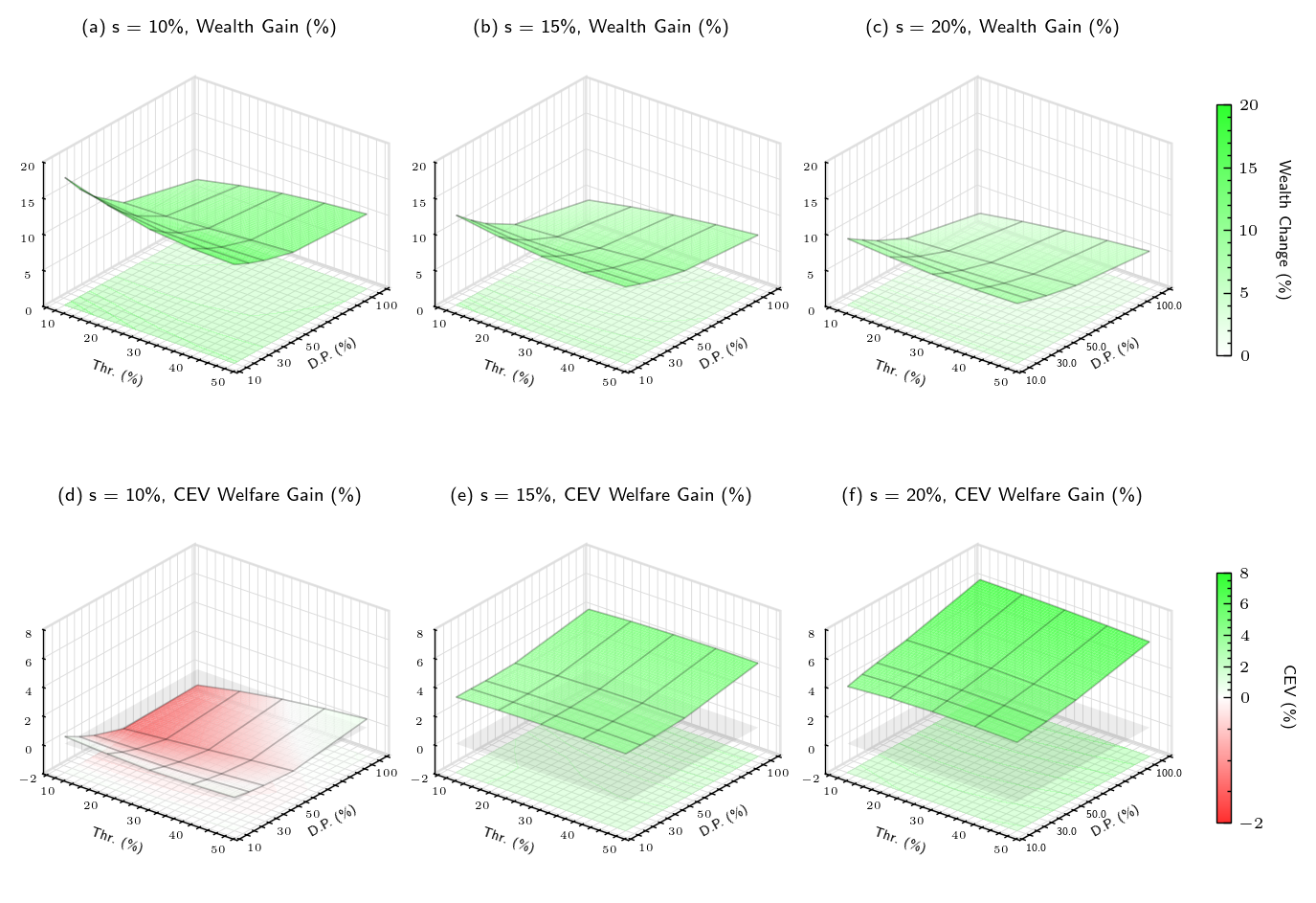}
    \caption{\textbf{Optimal Life-Cycle Homeowner Strategy}}
    \justifying{\footnotesize\noindent This figure presents terminal-wealth gains and lifetime consumption-equivalent variation (CEV) gains for single-home ownership strategies relative to a saving-rate-matched all-equity renter benchmark. Panels (a)--(c) report percentage gains in mean wealth at death for saving rates $s = 10\%$, $15\%$, and $20\%$. Panels (d)--(f) report the corresponding percentage CEV gains. In each panel, ``D.P. (\%)'' denotes the down payment ratio and ``Thr. (\%)'' denotes the purchase threshold that triggers home purchase. The vertical axis reports percentage gains relative to the matched renter benchmark.
    }
    \label{fig:homeownership_strategy_surface}
\end{figure}

Homeowners' wealth gains decrease monotonically with the saving rate, while consumption-equivalent gains rise with the saving rate. Both measures are positive across all strategy combinations at the 15\% and 20\% saving rates. The welfare gains reflect higher retirement consumption and larger bequests, despite consumption declines during the mortgage-servicing years. In addition to raising wealth and welfare on average, access to homeownership reduces wealth inequality at retirement, as measured by the Gini coefficient. The results are robust to region-specific samples, which yield similar life-cycle performance patterns. Figure~\ref{fig:homeownership_strategy_surface} summarizes these findings by visualizing the strategy grids for terminal wealth and consumption-equivalent gains, highlighting that leverage increases terminal wealth, whereas larger down payments increase consumption-equivalent gains.



We also observe substantial heterogeneity in the economic effects of homeownership. Low-down-payment strategies deliver the largest terminal-wealth gains, while delaying purchase allows households to accumulate a larger financial buffer that eases the consumption burden of mortgage payments, raising consumption-equivalent gains at the cost of modestly lower terminal-wealth gains. Gains also differ sharply by income. Low-income homeowners reduce consumption more sharply to meet mortgage obligations, consistent with evidence that the consumption of poorer, more levered households is highly sensitive to housing wealth \citep{mian2013household}. Many ultimately sell their homes to finance consumption, which further lowers their consumption-equivalent gains. The welfare benefits of homeownership are therefore concentrated among high-income households.

Market timing further shapes outcomes. Home purchases during a booming real estate market at the lowest mortgage rate lead to the greatest wealth and consumption-equivalent gains. Our interpretation is that in the long run, high-housing-price environments coincide with stronger macroeconomic conditions and better household income realizations, allowing households to service mortgages more safely and to build equity on a larger asset base. We find that low interest rates compress lifetime mortgage costs, preserve   balance-sheet capacity before retirement, and support smoother consumption. This heterogeneity challenges the one-size-fits-all advice against homeownership \citep{cocharne2022portfolios} and highlights the role of individual circumstances and market conditions in housing decisions.

Overall, homeownership preserves wealth, enhances welfare, protects households from the downside risks of their wealth portfolios, and mitigates wealth inequality. We propose four mechanisms behind this homeownership premium. First, housing reduces portfolio risk. As a bond-like asset with a stable implicit dividend of housing services, it lowers overall portfolio volatility and provides diversification benefits. We show that the average consumption-equivalent gain relative to the benchmark at the 15\% saving rate increases with risk aversion from approximately 1.0\% at $\delta=3$ to 3.4\% at the baseline value of $\delta=5$, and reaches 4.6\% at $\delta=10$. Homeowners in our simulations also face smooth maintenance costs alongside their mortgage payments, whereas renters face volatile rent-to-income ratios \citep{sinai2005owner}.

\begin{figure}[H]
    \centering
    \includegraphics[width=\linewidth]{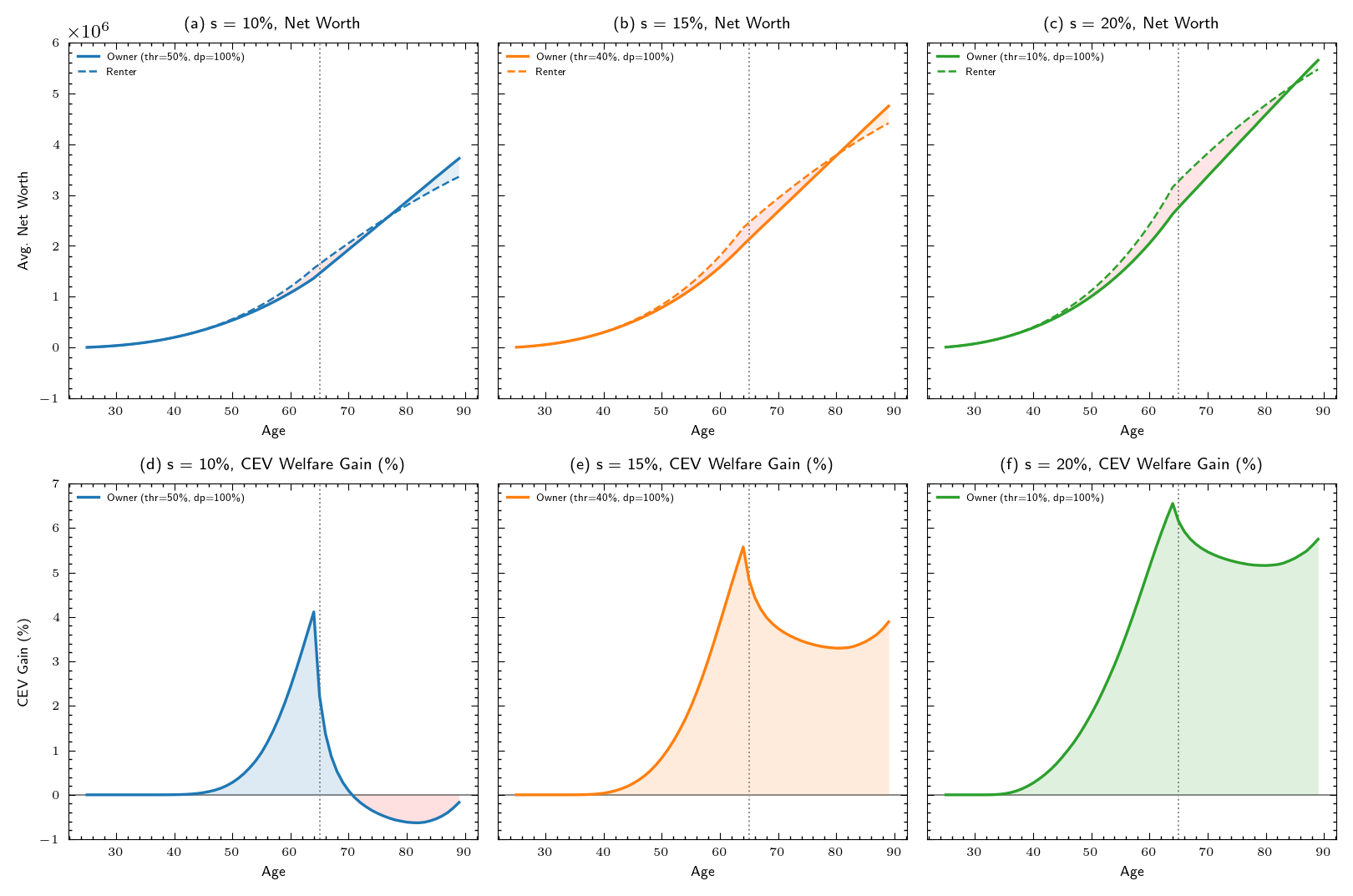}
    \caption{\textbf{Household Age Profile over the Life Cycle}}
    \justifying{\footnotesize\noindent This figure presents selected strategies' age profiles of household wealth and lifetime consumption-equivalent gains for homeowners and renters across saving rates. Wealth is measured in millions of 2024 U.S. dollars. Panels (a)--(c) report age profiles of the mean wealth across saving rates $s = 10\%$, $15\%$, and $20\%$. Panels (d)--(f) report the corresponding age profiles of percentage consumption-equivalent variation (CEV) gains. The vertical axis reports percentage gains relative to the matched renter benchmark. In each panel, ``D.P. (\%)'' denotes the down payment ratio and ``Thr. (\%)'' denotes the purchase threshold that triggers home purchase. The reported age profiles are based on the best-performing homeownership strategies in terms of lifetime consumption-equivalent gains at each saving rate. The best-performing strategies are: (i) 100\% down payment and 50\% purchase threshold at the 10\% saving rate, (ii) 100\% down payment and 40\% purchase threshold at the 15\% saving rate, and (iii) 100\% down payment and 10\% purchase threshold at the 20\% saving rate.
    } 
    \label{fig:age_profile_best_welfare}
\end{figure}

Second, homeownership induces intertemporal reallocation of wealth and welfare, resulting in gains later in life. During working years, mortgage servicing and housing investment crowd out liquid assets, so homeowners accumulate less financial wealth than renters \citep{becker2010outstanding, cocco2005housing, vestman2019limited, yao2004optimal}. However, as housing equity builds and is later converted into retirement consumption or bequests, wealth and consumption-equivalent gains rise significantly over time. Figure~\ref{fig:age_profile_best_welfare} illustrates this transition. At the 15\% saving rate, homeowner wealth initially lags but converges after retirement, while consumption-equivalent gains rise from near zero early in life to about 5.6\% at retirement and remain between 3.3\% and 3.9\% thereafter. The pattern is stronger at the 20\% saving rate, peaking at about 6.6\%.


Third, leverage trades consumption for earlier ownership. At the baseline 15\% saving rate, cash-purchase strategies generate consumption-equivalent gains of 4.17\% to 4.47\%, whereas mortgage-financed strategies yield 3.22\% to 3.55\%, as debt service compresses consumption during repayment. Therefore, cash-purchase strategies deliver about 1 percentage point higher consumption-equivalent gains than mortgage-financed strategies. Leverage also moves purchases forward by roughly a decade: Households making a 10\% down payment buy at a mean age of 33.6, versus 43.4 for cash buyers. In other words, delaying home purchase by a decade earns the household 1 percentage point in consumption-equivalent terms, while the same 1 percentage point in consumption-equivalent terms becomes the price of accessing housing services ten years earlier.


Fourth, consistent with the literature, homeownership reduces household liquidity in our simulation. At retirement, homeowners hold 26.1\% to 38.4\% less financial wealth than matched renters across the baseline strategy grid, leaving a thinner buffer against expenditure shocks. However, the gap substantially narrows to 9.3\% to 17.1\% by the end of life as households sell their homes or draw down home equity.


The rest of the paper proceeds as follows. Section~\ref{sec:related_literature} discusses the related literature and our contributions. Section~\ref{sec:life_cycle_model_simulation} describes the data and model setup. Section~\ref{sec:main_results} presents the quantitative results. Section~\ref{sec:mechanisms} examines the economic mechanisms behind the homeownership premium. Section~\ref{sec:second_home_ownership} examines second-home ownership. Section~\ref{sec:conclusion} concludes.

\section{Related Literature}
\label{sec:related_literature}

\noindent A long-standing question in household finance is whether homeownership is a sound investment for the typical household. Answering it requires evidence on the long-run behavior of housing returns, a life-cycle framework that captures how households actually finance and hold a home, and a realistic account of the risk borne on an individual property. Our paper draws on three strands of literature corresponding to these components and integrates their insights in a stationary block-bootstrap simulation of the household life cycle spanning 150 years of data from 16 countries.

The first strand evaluates housing through long-run aggregate price indices. \citet{shiller2001} documents that real U.S. home prices exhibit no continuous upward trend since 1890 \citep{shillerdata}, and \citet{eichholtz1997long} shows, using 345 years of data on Amsterdam's Herengracht, that real house prices can grow as slowly as 0.2\% per year. Meanwhile, \citet{choi2022popular} reports that popular personal finance books broadly advise against housing as an investment, and \citet{cocharne2022portfolios} argues that housing compares unfavorably with equity over long horizons. These studies provide essential benchmarks, but they capture only one dimension of the homeownership decision. Households hold a home for horizons far shorter than the centuries underlying these indices, and they benefit from leverage and imputed rental savings that price indices do not reflect.

Our paper translates this long-run price evidence into household-level outcomes. Simulating a full life-cycle balance sheet, we account for the holding horizon, leverage, and imputed rental savings that aggregate price indices omit; we thereby map long-run price dynamics onto the investment outcomes that matter for the rent-versus-own decision.

A second strand studies homeownership from the household's perspective. \citet{sinai2005owner} use Current Population Survey data from 1990 to 1999 to show that homeownership rates rise with a household's expected tenure horizon and the hedging benefit of owning against rent risk, and \citet{yao2004optimal} calibrate a life-cycle portfolio-choice model to Panel Study of Income Dynamics wealth supplements collected in five irregular waves between 1984 and 2001. These studies establish the household-level determinants of ownership, but their empirical foundations rest on short windows and infrequent balance-sheet snapshots, leaving open whether their conclusions hold over the multidecade horizons that housing decisions span.

Beyond these data limitations, the models themselves simplify the dependence structure of returns to manage the curse of dimensionality.  Stock return innovations are commonly modeled as independently and identically distributed (i.i.d.) \citep{gomes2005optimal}, a restriction \citet{yao2004optimal} extend to housing returns; they additionally parameterize the stock-housing relationship as a constant contemporaneous correlation and fix annual rent as a constant fraction of the property's value. \citet{cocco2005housing} treats house price shocks as uncorrelated with stock returns and holds the mortgage rate constant regardless of credit market conditions. These simplifications are necessary for tractability, but they rule out the serial dependence, time-varying comovement, and joint dynamics of prices, rents, and interest rates that the historical record displays.

Our approach addresses both limitations. Whereas \citet{duarte2022simple} confront the curse of dimensionality by solving richer dynamic programs with machine learning, we relax the parametric restrictions imposed by both dynamic programming and short panel data. Following \citet{politis1994stantionary} and \citet{anarkulova2023beyond}, we simulate life-cycle outcomes with a stationary block bootstrap that preserves the serial dependence and long-run comovement of prices, rents, and interest rates without imposing a parametric dependence structure. Combining this simulation with the income process of \citet{guvenen2021data} and 150 years of historical data across 16 countries captures multiyear autocorrelation and regime propagation over horizons that match the housing decisions households actually face. The breadth of the sample also addresses the generalizability and lucky-market concerns raised by \citet{anarkulova2022stocklongrun} and \citet{vanbinsbergen2025lucky}.

A third strand measures the risk of housing returns. Much of the life-cycle housing literature relies on national, regional, or metropolitan price indices, which capture common price movements but abstract from the distribution of realized household returns:  \citet{cocco2005housing} calibrates annual housing-return volatility at 6.2\%, and \citet{sinai2005owner} and \citet{yao2004optimal} similarly model aggregate or local market risk only. Recent granular evidence indicates that this abstraction is quantitatively important. Using property-level repeat-sales data, \citet{giacoletti2021idiosyncratic} finds that households face annualized housing-return risk of approximately 15.6\%, much of which is idiosyncratic and persists even at a five-year horizon (See also \cite{flavin2002owner}). \citet{bourassa2009house} similarly document substantial within-market heterogeneity in housing price appreciation across otherwise similar properties.

Our simulation closes this gap. Because we simulate a life-cycle household panel using real historical data rather than a parametric return process, the simulation naturally captures the risk a household faces on its own property: It delivers a total housing-return volatility of 15.3\%, closely matching the realized household-level risk documented by \citet{giacoletti2021idiosyncratic}. This allows us to quantify how household-level housing-return risk, alongside long-run aggregate risk, shapes the rent-versus-own decision, complementing the rent-hedging channel emphasized by \citet{sinai2005owner} with an explicit account of labor income risk.

Overall,   existing studies have advanced the field's  understanding of optimal household portfolio allocation along three distinct fronts, but no study has joined  the separate strands. By evaluating homeownership within a single framework that combines long-run cross-country return data, a flexible dependence structure, realistic household balance-sheet dynamics, and household-level housing-return risk, we provide a unified basis on which the rent-versus-own decision can be assessed.

\section{Life-Cycle Model Simulation}
\label{sec:life_cycle_model_simulation}

\noindent In this section, we describe our approach to simulating the life cycle of households. We first  introduce our data sources and discuss the simulated household panel. We then introduce the stationary block-bootstrap simulation and describe the economy, including household preferences, income dynamics, and  markets. At the end of this section, we discuss the investment strategies and the benchmarking specification.

\subsection{Data}
\label{subsec:data}
\noindent We adopt \citeauthor{jorda2019rate}'s \citeyearpar{jorda2019rate} long-run asset return data   for our price series. These data incorporate \citeauthor{katharina2017noprice}'s \citeyearpar{katharina2017noprice} historical housing market series, enabling us to study life-cycle portfolio allocation and the homeownership decision from a long-run perspective. Our choice of data source follows a growing literature that uses long-run macrofinancial series \citep{krishnamurthy2025, muir2017financial}.

The database spans from 1870 to 2020 across 18 high-income countries: Australia, Belgium, Canada, Denmark, Finland, France, Germany, Ireland, Italy, Japan, the Netherlands, Norway, Portugal, Spain, Sweden, Switzerland, the U.K., and the U.S. We  exclude Canada and Ireland for lack of data availability. The database provides macroeconomic series, such as government bills,   10-year bonds, stock market returns, dividends, housing capital gains,  rental yields, housing price indices, labor income growth, and inflation. Housing returns therefore come from a source that separates capital appreciation from rental income, which matters because the two behave differently over the long run. Although coverage varies across series, assets, and countries, each series is continuous once it enters the sample. 

For each economic path, the economic conditions are drawn from these series by a stationary block bootstrap, so a household experiences runs of consecutive years rather than independent draws. Our main results use the global data, with American household decision rules, which mitigates the concern that results driven by a single country's markets reflect luck rather than repeatable patterns \citep{anarkulova2022stocklongrun, vanbinsbergen2025lucky}; results based on regional subsamples are reported in Section \ref{subsec:regional_evidence}. Labor income follows   \citeauthor{guvenen2021data}'s \citeyearpar{guvenen2021data} process, which is estimated from U.S. panel data and captures heterogeneity in income growth across households. The income process is introduced in the next subsection.


\begin{table}[htbp]
    \justifying{\footnotesize\noindent
    This table reports summary statistics of the economic variables from the block-bootstrap-simulated household panel. All return, yield, interest rate, and inflation series are annual nominal rates expressed as percentages. Long-term bond return is the total return on 10-year government bonds. Housing return combines capital appreciation and rental yield, with the subseries reported separately. The inflation rate is derived from the consumer price index. Labor income growth is the annual percentage change in simulated labor income. HPIR is the house-price-to-labor-income ratio in levels.
    \vspace{4pt}}

    \centering
    \caption{\textbf{Summary Statistics}}
    \footnotesize
    \begin{tabularx}{\textwidth}{p{0.28\textwidth}YYYYYYY}
        \toprule
        & \mc{Mean} & \mc{Std. Dev.} & \mc{10\%} & \mc{25\%} & \mc{50\%} & \mc{75\%} & \mc{90\%} \\
        \midrule
        Domestic Equity Return      & 11.25 & 17.71 &  -9.60 &  2.36 & 10.68 & 19.66 & 36.40 \\
        International Equity Return & 10.58 & 14.01 &  -4.55 &  2.15 & 11.29 & 19.12 & 27.57 \\
        Long-Term Bond Return       &  5.97 &  8.22 &  -1.69 &  1.28 &  4.49 &  9.52 & 16.12 \\
        Housing Capital Gain        &  7.29 & 15.37 &  -1.29 &  2.53 &  5.29 & 10.17 & 14.87 \\
        Housing Rental Yield        &  4.03 &  1.20 &   2.24 &  3.38 &  3.99 &  4.61 &  5.23 \\
        Housing Return              & 11.31 & 15.28 &   2.55 &  6.74 &  9.20 & 14.10 & 19.56 \\
        Inflation Rate              &  4.03 &  5.08 &  -1.10 &  1.57 &  2.94 &  6.74 & 10.13 \\
        Labor Income Growth         &  0.67 & 28.20 & -26.52 & -4.94 &  0.00 &  7.70 & 25.36 \\
        HPIR                        &  4.56 &  1.37 &   3.40 &  3.64 &  4.03 &  4.79 &  6.96 \\
        \bottomrule
    \end{tabularx}
    \label{tab:return_summary_statistics}

\end{table}

Table~\ref{tab:return_summary_statistics} reports the summary statistics of the main economic variables across our block-bootstrap panel of 1~million simulated household life cycles. The international equity return is constructed as the equal-weighted index of all countries' equity total returns. Housing delivers equity-like average returns once rental income is included, but with a milder downside. Housing capital gains average 7.29\%, and adding rental yield raises the total housing return to 11.31\%, close to the 11.25\% average domestic equity return and consistent with the by-country time-series summary statistics in \citet{jorda2019rate}. 
Meanwhile, the lower tail of housing returns is much less severe than that of equity returns: The 10th percentile of total housing returns is 2.55\%, compared with $-9.60$\% for domestic equity. This return profile motivates the portfolio-level risk-reduction channel examined in Section \ref{sec:mechanisms}.

The remaining variables characterize the broader economic environment faced by households. Bond returns are lower and less volatile than equity returns, rental yields are stable relative to house price appreciation, and labor income growth is highly dispersed, reflecting substantial income risk in the simulated life-cycle panel. Finally, the house-price-to-labor-income ratio (HPIR) varies widely across countries and periods, so simulated households face substantially different affordability conditions at the time of their housing decisions.

\subsection{Simulation Design}
\label{subsec:simulation}
\noindent To preserve the serial and cross-market dependence in the data, we implement the stationary block-bootstrap approach following \citet{politis1994stantionary}. Each household is first randomly assigned to a country. We then draw blocks of consecutive years from that country's series until the full life cycle is covered, with block lengths following a geometric distribution with mean $T_s = 10$ years. Unlike a dynamic-programming approach, which requires specifying the joint stochastic processes in closed functional form parametrically, our simulation preserves the multivariate dependence structure without imposing distributional assumptions.

Each household (consisting of two members) is born at age 25, retires at age 65, and has a maximum age of 120. For each household, we separately simulate its members' labor income processes \citep{guvenen2021data} and impose the gender-based mortality risk from the Social Security Administration period life table \citep{ssa2024lifetable}. In other words, the  members face distinct labor income risks and longevity risks. Repeating this procedure yields a panel of 1~million simulated household life cycles.

All households start with zero wealth and make life-cycle investment decisions each period according to prespecified rules. Important decisions include the household's housing costs, discretionary savings, consumption, stock investment, decisions of home purchase and sale, and reverse mortgage decisions. We aggregate outcomes across different life-cycle strategies to evaluate the economic effects of homeownership.

\subsection{The Economy}
\label{subsec:the_economy}
\noindent In this subsection, we describe the economic environment in which households live and the decision rules they follow. A strategy \(\pi_j\) specifies the household's rules for consumption, liquid-asset investment, and housing investment. Household preferences are defined over equivalence-scale-adjusted consumption and terminal wealth \citep{duarte2022simple,denardi2010why}. Under strategy \(\pi_j\), household \(i\)'s lifetime utility is
\begin{align}
    \label{eq: utility}
    V_i(\pi_j)
    =
    \sum_{t=0}^{T_i}
    \beta^t
    \frac{
    \left(C_{it}(\pi_j)/\sqrt{N_{it}}\right)^{1-\delta}
    }{1-\delta}
    +
    \beta^{T_i} a_q
    \frac{
    \left(W_{iT_i}(\pi_j)+b_q\right)^{1-\delta}
    }{1-\delta}.
\end{align}
Here, \(C_{it}(\pi_j)\) denotes consumption, \(N_{it}\) is the number of surviving household members, and \(W_{iT_i}(\pi_j)\) is terminal wealth at the household's date of death \(T_i\). The parameter \(\beta\) is the discount factor, \(\delta\) is the coefficient of relative risk aversion, and \(a_q\) and \(b_q\) govern the bequest motive. We evaluate strategies by the expectation of \(V_i(\pi_j)\) across simulated life cycles.

Housing services do not enter the utility function directly. Both owners and renters consume housing and differ only in its financial consequences: Renters pay market rents, while owners bear the costs and returns of the housing asset. We thus evaluate homeownership purely as an investment, setting any nonpecuniary premium from owning to zero.\footnote{Survey and revealed-preference evidence suggests a positive ownership premium. See, for example,  \citet{sinai2005owner}. Our specification takes the zero-premium lower bound.} Any finding in favor of homeownership is therefore conservative.



\subsubsection{Income Dynamics}
\label{subsec:income_dynamics}
\justifying 
\paragraph{Labor Income} 
Within each working household, the labor income processes of the two spouses are assumed to be independent, and household labor income is the sum of the spouses' incomes. Following \citet{guvenen2021data}, the labor income $Y_{it}$ of each individual during working life is given by
\begin{align}
\label{eq: labor}
    Y_{it}=&(1-\gamma_{it})e^{g(t) + \alpha_i+\beta_{i}t+z_{it}+\varepsilon_{it}},
\end{align}
where $t=(\text{age}-24)/10$ is the normalized age, and $g(t)$ is a quadratic function of $t$. The pair \((\alpha_i, \beta_i)\) is randomly drawn from a bivariate normal distribution with zero mean and a covariance matrix estimated by \citet{guvenen2021data}. The variable \(\alpha_i\) captures ex ante heterogeneity in earnings levels, and \(\beta_i\) captures ex ante heterogeneity in earnings growth. The persistent component $z_{it}$ follows an AR(1) process,
\begin{align}
   z_{it} =& \lambda_z z_{it-1} + \nu_{it}, z_0\sim N(0,\sigma^2_{z_0}).
\end{align}
The innovation $\nu_{it}$ is drawn from the mixture of two normal distributions,
\begin{align}
    \nu_{it}\sim& \begin{cases}
 N(\mu_{\nu1},\sigma_{\nu1}^{2}) & \text{with probability } p_{\nu}.\\
  N(\mu_{\nu2},\sigma_{\nu2}^{2}) & \text{with probability } 1-p_{\nu}.
    \end{cases}
\end{align}
The transitory shock $\varepsilon_{it}$  follows a mixture normal distribution:
\begin{align}
 \varepsilon_{it}\sim& \begin{cases}
        N(\mu_{\epsilon1},\sigma_{\varepsilon1}^{2}) & \text{with probability } p_{\varepsilon}\\
        N(\mu_{\epsilon2},\sigma_{\varepsilon2}^{2}) & \text{with probability } 1-p_{\varepsilon}
    \end{cases}
\end{align}
Finally, the unemployment duration $\gamma_{it}$ follows
\begin{align}
      \gamma_{it} \sim &\begin{cases}
        0 & \text{with probability } 1-p_v(t,z_{it})\\
        \min(1, \exp(\lambda_{\gamma})) & \text{with probability } p_v(t,z_{it})
    \end{cases}\\
    p_v(t,z_{it}) =& \frac{1}{1+\exp(-\xi_{it})},\quad \xi_{it} = a_{\xi}+b_{\xi}t + c_{\xi}z_{it}+ d_{\xi}tz_{it}
\end{align}
where the probability of unemployment, $p_{v}$, depends on age and the persistent component so that job loss is more likely for workers with depressed persistent earnings.


\justifying 
\paragraph{Retirement Income} 
Retirees in our economy have two income sources: SSRBs and withdrawals from liquid wealth. SSRBs  are specified as a constant fraction $\lambda$ of the average labor income over the last five working periods:
\begin{align}
    Y_{it} = \lambda \bar{Y}_{iR-5:R-1},\quad t\geq R
\end{align}
The retirement period $R=40$, corresponding to the retirement age $65$. 

In addition to SSRBs, households withdraw from liquid wealth to finance retirement consumption. Each period, the target withdrawal is the greater of a fixed fraction of current liquid wealth and the prior period's withdrawal, so that withdrawals never decline in retirement.

\justifying  
\paragraph{SSI}
We also incorporate a government-provided consumption floor into the income process, which we refer to as SSI. SSI supplements household income to exactly the minimum consumption level and is available only to renters whose resources are insufficient to finance that level; the supplement ceases once labor income or retirement benefits lift
household resources above the floor. We term the exhaustion of economic resources financial ruin \citep{anarkulova2023beyond}.

Homeowners are not eligible for SSI while they retain their home. A working-age homeowner who cannot meet the consumption floor must sell the home and become a renter. If a retiree's SSRBs and withdrawals do not cover the consumption floor, the household draws on home equity through a reverse mortgage or a sale of the home and then exhausts its remaining financial resources to maintain consumption. Once all resources are depleted, the household receives the SSI consumption floor.\footnote{We apply the same floor to working-age households because the transfer system replaces SSI at those ages with food assistance, refundable tax credits, unemployment insurance, and health care support, which together deliver no less assistance. Representing these programs jointly as a single floor across the life cycle follows \citet{hubbard1995precautionary}.} The supplement ceases when labor income or retirement benefits lift household resources above the floor.

\justifying 
\paragraph{Rental Property Income} 
Our main results consider ownership of a single, owner-occupied home. In an extension at the end of the paper, we allow households with sufficient liquidity to purchase a second home, which generates rental income according to a simulated rental yield.

\subsubsection{Markets}\label{sec:markets}
\justifying 
\paragraph{Financial Markets} 
Households have access to stock and bond markets, a housing market, and credit markets offering a 30-year fixed-rate mortgage and a fixed-rate reverse mortgage (FRRM). In period \(t\), household \(i\) faces household-specific returns and rates \(\{r^s_{it}, r^b_{it},
r^h_{it}, r^c_{it}, r^{hpir}_{it}, r^d_{it}, y^m_{it}, y^{rm}_{it}\}\), corresponding to stocks, bonds, housing capital, rental yields, the HPIR, inflation, the mortgage rate, and the reverse mortgage rate, respectively. The joint distribution of returns, the HPIR, and inflation is given by
\begin{align}
    &\{(r^s_{it}, r^b_{it}, r^h_{it},r^c_{it},r^{hpir}_{it},r^d_{it})\}^{T}_{t=1}\sim \tilde{F},\\
    &y^m_{it} = y^b_{it} + \text{mortgage spread},\\
    \label{eq: rates}
    &y^{rm}_{it} = y^b_{it} + \text{reverse mortgage spread},
\end{align}
where $\tilde{F}$ denotes the distribution from the simulation and $y^b_{it}$ denotes the yield on bonds.

In the simulation, we consider portfolios with different asset allocations, taking the internationally diversified all-equity as the benchmark for comparison \citep{anarkulova2023beyond}.\footnote{Household savings are invested entirely in equities, with one-third allocated to domestic equity and two-thirds to an equal-weighted international equity portfolio constructed across all countries in the Macrohistory Database sample.} We abstract from taxes on stock returns. This assumption is conservative for our quantitative results: Tax-deferred retirement accounts postpone, rather than eliminating, taxation since withdrawals are taxable. By treating stock returns as untaxed, the model makes financial saving more attractive than it would be under the tax system. Therefore, if homeownership increases life-cycle wealth accumulation, our estimates understate the economic magnitude of this effect.

\paragraph{Housing Market} 
All households enter the model as renters. While renting, they pay the rental cost $r^c_{it}P^h_{it}$, where $P^h_{it}$ is the price of one unit of housing capital. Households assigned to homeownership strategies transition from renting to owning once they meet the home-purchase eligibility criteria. To capture housing market frictions, we assume   households pay a transaction cost equal to a fraction \(c_h\) of the house value whenever they buy or sell a home, reflecting legal fees, broker commissions, and loan origination fees.

At household entry, the home price is set to 4.06 times the household's labor income.\footnote{The HPIR anchor is set   using the median sales price of houses sold \citep{census2024mspus} and median household income \citep{census2024mehoin} from 1990, the year the nominal house price index is normalized to 100 by \citet{jorda2019rate}. After entry, as  home price and labor income change based on the simulation, the HPIR changes accordingly.} Thereafter, the home price evolves according to the simulated housing capital return \(r^h_{it}\), so the price-to-income ratio varies endogenously over the life cycle. Homeowners pay a per-period maintenance cost $c_{m}H_{it}$, where $H_{it}$ is the house value and \(c_m\) measures the maintenance cost per unit of total property value. These costs include property taxes, homeowners' association fees, homeowners' insurance, and routine maintenance and repairs. Homeowners who lease a second home with housing capital $H^v_{it}$ receive rental yield $r^c_{it}H^v_{it}$. 



\paragraph{Home Purchase}
We consider two purchase approaches: cash purchase and mortgage financing. Regardless of the purchase approach, the home purchase requires that the current household labor income must exceed the sum of the minimum consumption floor, the required mortgage payment, and the annual maintenance cost of the property. 

We further assume that a household can finance the home purchase when three additional conditions are satisfied. First, household labor income net of the minimum consumption floor must cover at least three times the mortgage payment plus the annual maintenance cost. We term this the payment-to-income requirement. It mirrors the Federal Housing Administration's  standard ``front-end ratio'' of 31\% \citep{hud2024handbook} and the 36\% debt-to-income ratio limit used by Fannie Mae and Freddie Mac \citep{fanniemae2024selling}. It ensures  the household can service the debt without immediate financial distress \citep{li2007lifecycle}. Second, financial wealth must cover the required down payment, the additional purchase threshold, and the transaction costs. In particular, we consider different combinations of down payments (10\%, 20\%, 30\%, 50\%, and 100\% of the home value) and purchase thresholds (10\%, 20\%, 30\%, 40\%, and 50\% of the home value). Third, the household must have no prior default on its mortgage. 

These eligibility conditions reflect the joint importance of liquid assets and down payment constraints in shaping the homeownership decision \citep{yao2004optimal,cocco2005housing}. The purchase threshold, a buffer above the down payment, captures the empirical pattern, documented by \citet{flavin2002owner}, in which the house-price-to-net-worth ratio decreases from 3.51 for the youngest homeowners to 0.65 for the oldest. Cash purchase is also considered in our simulation. In the extension analysis, a household can finance a second home purchase once the first mortgage is fully paid off.



\justifying 
\paragraph{Reverse Mortgage}
Senior households are more likely to be equity rich but cash poor, and they face a higher risk of financial distress. We therefore consider the reverse mortgage as a way for retired households to cash out home equity and maintain consumption. The FRRM allows the household to borrow against home equity after retirement, and it does not require repayment until termination--- when the household dies, sells the home, or no longer uses it as its primary residence. The FRRM is costly: Borrowers pay an origination cost at initiation, an annual insurance premium, and an annual service fee. The borrower receives a lump-sum payment at  origination, and the FRRM balance grows at a rate equal to the prevailing bond rate plus the reverse mortgage spread, which combines the origination amortization, the annual mortgage insurance premium, and the annual service fee. The households can repay the FRRM out of housing equity at termination, and any balance over the home value is insured. 

The FRRM also constrains how much households can borrow against home equity. We follow the Federal Housing Administration's Home Equity Conversion Mortgage   program in setting the percentage of home value that households can borrow, which the program defines as the principal limit factor.\footnote{See the precomputed principal limit factor tables in \citet{hud2024hecm}.} This  factor is jointly determined by the age of the youngest borrower and the interest rate, which is the bond rate in our model.\footnote{In our simulation, individuals in a household have the same age.}

A retired household can secure a reverse mortgage when four conditions are satisfied. First, its combined income from SSRBs and liquid-asset withdrawal is sufficient to cover ongoing home maintenance costs so that the household can remain in the house. Second, the amount available under the principal limit factor exceeds any remaining mortgage balance, ensuring that positive net equity is available to borrow against. Third, its combined retirement income from SSRBs and the floor withdrawal is at or below the minimum consumption threshold. Fourth, the household is below age 95. When all conditions are met, the household borrows the maximum Home Equity Conversion Mortgage--eligible amount and does not repay until it liquidates the home. In the extension analysis with two houses, households take a reverse mortgage on the primary residence only  since the reverse mortgage market requires the collateral to be the borrower's primary residence.

\paragraph{Home Liquidation} 

Because housing is illiquid, households facing financial difficulties may be forced to liquidate their homes. We consider two liquidation rules. First, households liquidate their home when the current loan-to-value  ratio rises above 1.5 after loan origination.\footnote{Because households cannot borrow more than the house value at purchase, the loan-to-value ratio is at most 1 at origination; a ratio above 1.5 means the mortgage balance exceeds 150\% of the market value of the house.} Second, households liquidate their home when the minimum consumption level exceeds total available liquid resources---that is, when current-period labor income plus prior-period financial wealth plus any secondary-home rental income, less ongoing maintenance costs, is insufficient to meet the minimum consumption threshold \citep{denardi2010why,hubbard1995precautionary}. For households with two homes, the loan-to-value  rule applies home by home: Any  home with an  ratio above 1.5 is liquidated. Under the consumption rule, two-home households first liquidate the second home and become single-home owners; if resources remain insufficient, they then liquidate the primary residence.

Liquidation takes one of two forms: sale or default.\footnote{We do not model mortgage delinquency. Households either continue servicing their mortgage, sell the house, or default on the mortgage.} Households sell when the net sale proceeds cover the outstanding mortgage balance---that is, when $(1 - c_h) \times H_{it} \geq M_{it}$, where $M_{it}$ is the remaining mortgage balance, and default otherwise.

\subsubsection{Saving and Consumption}
\noindent Before retirement, households save a fixed fraction of net labor income (labor income minus fixed costs) and consume the rest. We examine saving rates of 10\%, 15\%, and 20\%, following the comprehensive household saving evidence in \citet{dynan2004rich} and the popular advice surveyed by \citet{choi2022popular}. We regard 15\% as the baseline saving rate and use the other two rates for robustness checks. The 15\% baseline is broadly consistent with the rate advocated by major financial institutions (e.g., \citealt{fidelity2024saving}; \citealt{troweprice2024saving}; \citealt{vanguard2024howamerica}; \citealt{schwab2025starting}).\footnote{Industry guidance identifies 15\%, inclusive of employer contributions, as the saving rate that preserves preretirement living standards. \citet{fidelity2024saving} and \citet{troweprice2024saving} advise saving at least 15\% of income, the former because 15\% accumulates ten times preretirement income by age 67 and with Social Security sustains preretirement consumption, the latter to fund a retirement of 30 years or more against longevity, inflation, and health-cost risk. \citet{vanguard2024howamerica} recommends a total contribution rate of 12\% to 15\% inclusive of employer contributions, and its annual survey of defined-contribution plan participants reports median 401(k) contribution rates of approximately 11\%, inclusive of employer match. Because this figure reflects employer-plan contributions alone and excludes other saving vehicles, total household saving can easily exceed this amount and become consistent with a saving target of 15\% or higher.}$^,$\footnote{Since our model abstracts from employer-sponsored pensions and Social Security, the saving rate in our framework captures all forms of household saving in a single measure, making it directly comparable to the comprehensive saving rate in \citet{dynan2004rich}.} Moreover, homeownership in our model is effectively limited to above-median earners since buyers must satisfy the eligibility conditions in Section~\ref{sec:markets}. For this group, \citet{huggett2000saving} report an empirical saving rate of about 13\% and a model-implied optimal rate of about 18\%. The 15\% baseline therefore lies at the center of the empirically observed and model-implied range for the homeownership-eligible population.

The saving-rate grid also serves an economic purpose beyond robustness. The 10\% case provides a conservative lower boundary that reveals when mortgage service suppresses working-life consumption enough to overturn the consumption-equivalent gain, whereas the 15\% and 20\% cases show when housing's wealth gains translate reliably into consumption-equivalent gains against the all-equity renter benchmark.

\subsubsection{Wealth Accumulation}  
\label{subsec:wealth_accumulation}

\noindent At each period $t$, household $i$ enters with accumulated real wealth $W_{it}$ and receives labor income $Y_{it,1}$ from the household head and, if the spouse is employed, $Y_{it,2}$ from the spouse. When applicable, the household receives other cash inflows $\varphi_{it}$ (including home liquidation proceeds, rental income from a second home, proceeds from mortgage and reverse mortgage borrowing, SSRBs, and SSI) and incurs expenditures $\phi_{it}$ (including mortgage payments, home transaction costs, housing maintenance for owners, and rent for renters).

The household chooses consumption $C_{it}$ and allocates a share $\alpha_{it}$ of the remaining wealth to stocks, with the rest invested in bonds. Real wealth then evolves as
\begin{equation}
    W_{i,t+1} = (1 + r^p_{it})\left( \frac{W_{it}}{1 + r^d_{it}} + Y_{it,1} + Y_{it,2} + \varphi_{it} - \phi_{it} - C_{it} \right),
\end{equation}
where $r^d_{it}$ is the inflation rate that converts nominal values into real terms  and $r^p_{it}$ is the portfolio return given by
\begin{equation}
    r^p_{it} = \alpha_{it}\, r^s_{it} + (1-\alpha_{it})\, r^b_{it}.
\end{equation}

\begin{table}[!htbp]
    \small
    \centering
    \caption{\textbf{Parameterization}}
    
    \justifying{\footnotesize\noindent This table presents the parameter settings of our simulation. Panel~A reports the preference and demographic parameters. Panel~B reports the labor income parameters, all from \citet{guvenen2021data}. The deterministic component $g(t)$ is quadratic in normalized age $t=(\text{age}-24)/10$ and the parameters are estimated using 2010 as the base year before rescaling to 2024 U.S. dollars. Panel~C reports the public-transfer parameters. Panel~D reports the housing and ownership parameters, where the price-index anchor at household entry is expressed as a house-price-to-labor-income ratio (HPIR). Panel~E reports the credit parameters, with spreads quoted over the bond rate drawn from the simulated economy. Panel~F reports the saving, withdrawal, and portfolio parameters, where the retirement withdrawal ratchets so the household draws the greater of 4\% of current financial wealth and its prior-period withdrawal. Currency values are 2024 U.S. dollars.\par\vspace{4pt}}
    \setlength{\tabcolsep}{3pt}\renewcommand{\arraystretch}{0.90}
    \footnotesize
    \noindent\begin{tabularx}{\textwidth}{lYlY}
    \toprule
    \multicolumn{4}{c}{Panel A: Preferences and Demographics} \\
    \midrule
    Parameter & Value & Parameter & Value \\
    \midrule
    Risk aversion $\delta$ & 5 & Time preference $\beta$ & 0.99 \\
    Bequest intensity $a_q$ & 2,360 & Bequest curvature $b_q$ & \$490,000 \\
    Life cycle & 25 to 120, retire at 65 & Equivalence scale & $\sqrt{2}$ to 1 \\
    Longevity & SSA Life Table 4c6 & & \\
    \midrule
    \multicolumn{4}{c}{Panel B: Labor Income} \\
    \midrule
    Parameter & Value & Parameter & Value \\
    \midrule
    Age profile $g(t)$ & $a_0=2.581$, $a_1=0.812$, $a_2=-0.185$ & Heterogeneity $\alpha,\beta$ & $\sigma_{\alpha}=0.3$, $\sigma_{\beta}=0.196$, $\rho=0.768$ \\
    Persistent shock $z_{it}$ & $\lambda_{\nu}=0.959$, $p_{\nu}=0.407$, $\sigma_{z_0}=0.714$, $\mu_{\nu}=(-0.085,0.058)$, $\sigma_{\nu}=(0.364,0.069)$ & Transitory shock $\epsilon_{it}$ & $p_{\epsilon}=0.13$, $\mu_{\epsilon}=(0.271,-0.040)$, $\sigma_{\epsilon}=(0.285,0.037)$ \\
    Nonemployment $\gamma_{it}$ & $\lambda_{\gamma}=0.01\%$, $a_{\xi}=-3.036$, $b_{\xi}=-0.917$, $c_{\xi}=-5.397$, $d_{\xi}=-4.442$ & & \\
    \midrule
    \multicolumn{4}{c}{Panel C: Public Transfers} \\
    \midrule
    Parameter & Value & Parameter & Value \\
    \midrule
    Replacement rate $\lambda$ & 0.65 of last 5-year average & Consumption floor & \$11,316 plus \$5,664 spouse \\
    \midrule
    \multicolumn{4}{c}{Panel D: Housing and Ownership} \\
    \midrule
    Parameter & Value & Parameter & Value \\
    \midrule
    Maintenance $c_m$ & 2\% of home value & Transaction cost $c_h$ & 2\% each way \\
    Payment-to-income & 1/3 & Liquidation trigger & Loan-to-value 1.5 \\
    Down payment & 10, 20, 30, 50, 100\% & Purchase threshold & 10, 20, 30, 40, 50\% \\
    Sale versus default & Sell if $(1-c_h)H \geq M$ & HPIR anchor at entry & 4.06 at household entry \\
    \midrule
    \multicolumn{4}{c}{Panel E: Credit} \\
    \midrule
    Parameter & Value & Parameter & Value \\
    \midrule
    Mortgage & 30-year fixed & Mortgage spread & 1.85\% \\
    Reverse mortgage & 3.35\% spread, 2\% origination, 0.5\% insurance, 0.5\% service & Principal limit factor & 0.41 to 0.64 of home value \\
    \midrule
    \multicolumn{4}{c}{Panel F: Saving, Withdrawal and Portfolio} \\
    \midrule
    Parameter & Value & Parameter & Value \\
    \midrule
    Saving rate & 10, 15, 20\%, baseline 15\% & Equity sleeve & 1/3 domestic, 2/3 international \\
    Retirement withdrawal & Greater of 4\% of wealth or prior withdrawal & & \\
    \bottomrule
    \end{tabularx}
    \label{tab:param_household}
\end{table}

\subsubsection{Parameterization}
\label{subsec:parameterization}
\noindent Table~\ref{tab:param_household} summarizes the model parameterization. We set the risk-aversion coefficient to $\delta=5$, a standard value in the life-cycle portfolio literature \citep{yao2004optimal,gomes2005optimal,koijen2010lifecycle,li2007lifecycle,cocco2005housing,campbell2003multivariate}. Following \citet{denardi2010why} and \citet{anarkulova2023beyond}, we set the bequest intensity to $a_q=2{,}360$ and the bequest curvature parameter to $b_q=\$490{,}000$ in 2024 U.S. dollars. The annual time-preference factor is $\beta=0.99$. Labor income is specified as in \citet{guvenen2021data}, scaled to match average log earnings in 2024 U.S. dollars. The minimum consumption level is set to \$11{,}316 for an individual, with an additional \$5{,}664 for a spouse, corresponding to the 2024 U.S. SSI federal benefit rates.\footnote{Unless otherwise noted, all dollar amounts are expressed in 2024 U.S. dollars.}

\begin{figure}[htbp]
    \centering
    \includegraphics[width=0.68\linewidth]{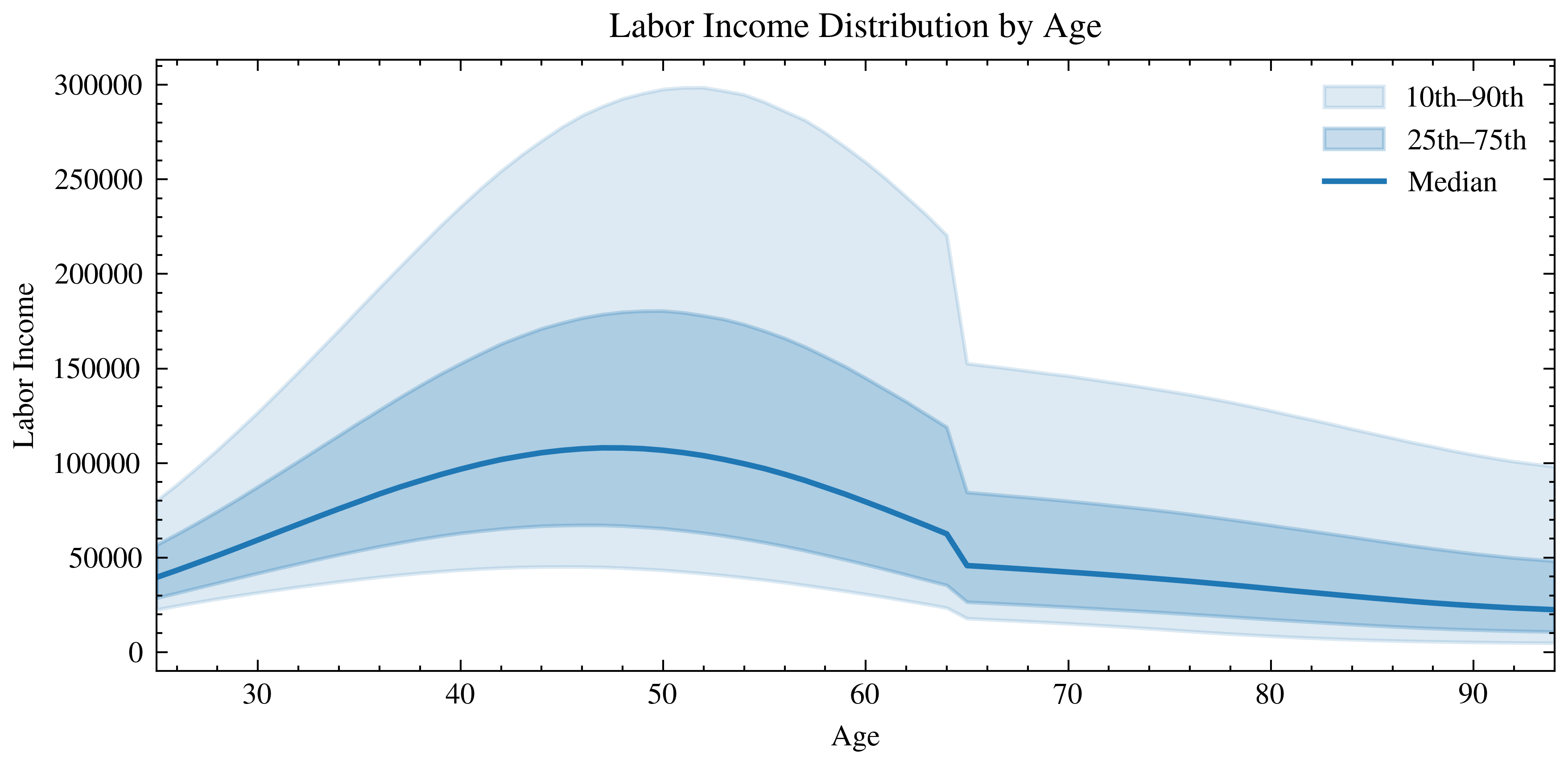}
    \caption{\textbf{Labor Income Fan Chart by Age}}
    \justifying{\footnotesize\noindent This figure presents the cross-sectional distribution of household labor income by age in our simulated panel. The labor income process is based on \citet{guvenen2021data}.}
    \label{fig:labor_income_fan_chart}
\end{figure}

Figure~\ref{fig:labor_income_fan_chart} shows the simulated cross-sectional distribution of labor income over the life cycle. Median income rises from about \$40{,}000 at age 25 to a peak near \$105{,}000 around ages 45 to 50, then declines modestly before dropping sharply at age 65, when labor income is replaced by SSRBs. The right-skewed dispersion during working years reflects the permanent and transitory shocks in the income process, and the resulting replacement gap at retirement motivates the range of saving rates we examine.

We set the mortgage spread to 1.85\% and the reverse mortgage spread to 3.35\%, estimated as the average spread in U.S. monthly data from January 2016 to May 2024. The reverse mortgage spread is inclusive of the annual service fee, and it is close to \citeauthor{nakajima2017reverse}'s \citeyearpar{nakajima2017reverse} figure. Because \citet{jorda2019rate} normalize the wage and housing price indices to 100 in 1990, the index level carries no economic units; therefore, we rescale the index to  4.06 in 1990 calibrated to U.S. data; from this anchor, home prices evolve through each household simulation according to the simulated housing capital-appreciation series. The payment-to-income ratio at purchase is set to 1/3. Following the literature, we impose an annual maintenance cost of 2\% of the home value \citep{cocco2020aging,yao2004optimal,cocco2005housing,nakajima2017reverse} and consider alternative values in the mechanism analysis of Section~\ref{sec:mechanisms}. The transaction cost is 2\% of the home value at both purchase and sale. Consistent with standard Home Equity Conversion Mortgage program terms, we model a 2\% origination cost at initiation, a 0.5\% annual mortgage insurance premium, and a 0.5\% annual service fee. The retirement withdrawal from the financial portfolios is set to the greater of 4\% of current financial wealth and the prior-period withdrawal \citep{bengen1994withdraw4pct,choi2022popular}.


\subsection{Investment Strategies} 
\label{subsec:investment_strategies}
\noindent For each simulated household, we fix the lifetime path of shocks and evaluate outcomes under every strategy, so cross-strategy differences are driven by the strategies themselves. Aggregating across households, we then compare wealth and welfare outcomes under homeownership against a set of investable liquid portfolio strategies.

The investable liquid portfolio strategies are constructed from three asset classes: domestic and international equity market total return indices, 10-year government bond returns, and short-term government bill returns. From these, we form four strategies: an all-equity portfolio, a balanced 60/40 equity-bond portfolio, a 100-minus-age rule, and a target-date fund (TDF).\footnote{We also computed all-bond and all-bill strategies but do not report them, as they underperform all reported strategies across our outcome measures.}

Specifically, each strategy involving equity has both an internationally diversified variant and a domestic-only variant. Following \citet{anarkulova2023beyond}, the internationally diversified stock sleeve holds one-third domestic equity and two-thirds international equity. The 60/40 strategy holds a constant 60\% in stock and 40\% in bond, rebalanced annually. The 100-minus-age strategy sets the stock allocation at 75\% at age 25 and reduces it linearly to 0\% at age 100.

The TDF follows Vanguard's glide path \citep{vanguardglidepath}. The equity share is 90\% through age 40, declines linearly to 60\% at age 60 and  50\% at age 65, then declines to a landing allocation of 30\% at age 72, which is held for the remainder of life. The non-equity share is invested in 10-year government bonds, with short-term government bills introduced from age 60. Vanguard reports phase boundaries and allocations rather than annual weights, so we interpolate linearly within each phase, implying equity-share declines of 1.5, 2.0, and 2.85 percentage points per year over ages 40--60, 60--65, and 65--72, respectively.


\subsection{Benchmarking} 
\label{subsec:benchmarking}
\noindent We focus on the internationally diversified all-equity strategy as the main benchmark for two reasons. First, homeownership and stock holdings occupy the same slot in the household portfolio, from the perspectives of both empirical evidence and life-cycle theory. Empirically, owner-occupied housing crowds out equity participation \citep{cocco2005housing,flavin2002owner,yao2004optimal,campbell2006household}.\footnote{\citet{cocco2005housing} documents that introducing owner-occupied housing reduces equity market participation from approximately 76\% to 33\% among homeowners.} Theoretically, because human capital functions as an implicit long-duration bond, the optimal financial portfolio for a household without housing tilts heavily toward equities \citep{cocco2005consumption,gomes2005optimal}, so renters without a housing substitute should hold equities aggressively, consistent with our benchmark. The all-equity strategy therefore provides the cleanest counterfactual for evaluating whether housing is a better wealth-creation vehicle than stocks, directly addressing \citeauthor{cocharne2022portfolios}'s \citeyearpar{cocharne2022portfolios} critique that housing is an illiquid and concentrated asset that underperforms equities.

Second, the all-equity strategy sets a high bar for homeownership to add value. \citet{anarkulova2023beyond} show that the all-equity strategy dominates common life-cycle alternatives in building retirement wealth and welfare. We confirm that it outperforms the alternative benchmarks in our setting (the 60/40, 100-minus-age, TDF, and [unreported] bond-only strategies) at a moderate risk-aversion coefficient, consistent with \citet{choi2025practical}, who show that all-equity is near optimal under constant relative risk aversion  utility with a risk-aversion coefficient of 4.\footnote{\citet{choi2025practical} report a consumption-equivalent loss of only 0.56\% relative to the optimum in \citet{cocco2005consumption}.}

Because we adopt the internationally diversified all-equity strategy as our main benchmark, households with access to homeownership invest their additional liquid wealth in the same strategy; for parsimony, we do not construct homeowner variants of each alternative liquid strategy. Our main results compare households by assigned strategy rather than by realized ownership. Households assigned a homeowner strategy that never satisfy the eligibility rules for purchase remain in the homeowner group; under the strategy's rules, they continue to hold the all-equity portfolio while waiting to qualify, so they follow the strategy throughout. This intention-to-treat design avoids selection on the endogenous realization of home purchase: Households that realize ownership are disproportionately those with favorable income paths, so conditioning on realized ownership would confound the value of the strategy with the luck of the paths that enable purchase.\footnote{The design parallels intention-to-treat analysis in clinical trials, which compares patients by treatment assignment rather than compliance because compliance itself is  selected.} Section~\ref{sec:mechanisms} examines the realized-buyer population directly.

\section{Quantitative Results}
\label{sec:main_results}
\noindent We evaluate simulation outcomes for households implementing different life-cycle strategies, with homeowner strategies defined over a grid of down-payment requirements and purchase eligibility thresholds. Each homeowner strategy is compared against a set of renter benchmarks that follow common financial-asset allocation rules, with the internationally diversified all-equity renter serving as the primary baseline, as discussed above. All comparisons hold the discretionary saving rule fixed across homeowners and renter strategies, ensuring that any observed advantage of homeownership is not driven by differential discretionary saving behavior. 

\FloatBarrier
\begin{table}[p]
    \justifying{\noindent \footnotesize This table reports outcomes from stationary block-bootstrap simulations of 1~million household life cycles. The homeowner and the matched renter face identical labor-income, asset-return, inflation, and survival paths and follow the same saving rule after fixed costs. The renter rents for life and invests only in stocks; the homeowner purchases once financial wealth covers the ``Down Payment'' plus an additional purchase ``Threshold,'' both as percentages of home value. Panels~A, C, and E report wealth gains at death for saving rates of 10\%, 15\%, and 20\%; Panels~B, D, and F report the corresponding consumption-equivalent variations (CEVs) at death. All outcomes are expressed relative to the saving-rate-matched all-equity renter benchmark. Wealth gains are percentage differences in mean wealth at death. CEV is the percentage difference between the homeowner and renter benchmark strategies in the constant consumption levels that replicate each strategy's realized lifetime utility, averaged across households. A positive (negative) CEV is a consumption-equivalent gain (loss). All values are in percent.\vspace{4pt}}\par
    \caption{\textbf{Life-Cycle Gains}}
    \centering
    \footnotesize
    \begin{tabularx}{\textwidth}{p{.15\linewidth}YYYYY}
        \toprule
        \multicolumn{6}{c}{Panel A: Wealth Gains at Death, Saving Rate\,=\,10\%} \\
        \midrule
        & \multicolumn{5}{c}{Down Payment} \\
        \cmidrule{2-6}
        \multicolumn{1}{c}{Threshold}
        & 10\% & 20\% & 30\% & 50\% & 100\% \\
        \midrule
        \multicolumn{1}{c}{10\%}&    17.72 &    14.99 &    12.78 &     9.63 &     7.10 \\
        \multicolumn{1}{c}{20\%}&    15.91 &    13.78 &    12.09 &     9.76 &     8.30 \\
        \multicolumn{1}{c}{30\%}&    14.56 &    13.00 &    11.76 &    10.10 &     9.22 \\
        \multicolumn{1}{c}{40\%}&    13.91 &    12.74 &    11.81 &    10.50 &     9.88 \\
        \multicolumn{1}{c}{50\%}&    13.69 &    12.74 &    11.96 &    10.82 &    10.37 \\
        \midrule
        \multicolumn{6}{c}{Panel B: CEVs at Death, Saving Rate\,=\,10\%} \\
        \midrule
        & \multicolumn{5}{c}{Down Payment} \\
        \cmidrule{2-6}
        \multicolumn{1}{c}{Threshold}
        & 10\% & 20\% & 30\% & 50\% & 100\% \\
        \midrule
        \multicolumn{1}{c}{10\%}&     0.50 &    -0.09 &    -0.57 &    -1.24 &    -1.09 \\
        \multicolumn{1}{c}{20\%}&     0.31 &    -0.14 &    -0.50 &    -0.98 &    -0.51 \\
        \multicolumn{1}{c}{30\%}&     0.18 &    -0.13 &    -0.39 &    -0.70 &    -0.05 \\
        \multicolumn{1}{c}{40\%}&     0.21 &    -0.03 &    -0.21 &    -0.41 &     0.30 \\
        \multicolumn{1}{c}{50\%}&     0.31 &     0.13 &    -0.01 &    -0.15 &     0.59 \\
        \midrule
        \multicolumn{6}{c}{Panel C: Wealth Gains at Death, Saving Rate\,=\,15\%} \\
        \midrule
        & \multicolumn{5}{c}{Down Payment} \\
        \cmidrule{2-6}
        \multicolumn{1}{c}{Threshold}
        & 10\% & 20\% & 30\% & 50\% & 100\% \\
        \midrule
        \multicolumn{1}{c}{10\%}&    12.48 &    10.65 &     9.04 &     6.61 &     4.25 \\
        \multicolumn{1}{c}{20\%}&    11.51 &    10.00 &     8.73 &     6.92 &     5.33 \\
        \multicolumn{1}{c}{30\%}&    10.82 &     9.66 &     8.72 &     7.36 &     6.21 \\
        \multicolumn{1}{c}{40\%}&    10.56 &     9.68 &     8.93 &     7.83 &     6.90 \\
        \multicolumn{1}{c}{50\%}&    10.59 &     9.85 &     9.22 &     8.23 &     7.46 \\
        \midrule
        \multicolumn{6}{c}{Panel D: CEVs at Death, Saving Rate\,=\,15\%} \\
        \midrule
        & \multicolumn{5}{c}{Down Payment} \\
        \cmidrule{2-6}
        \multicolumn{1}{c}{Threshold}
        & 10\% & 20\% & 30\% & 50\% & 100\% \\
        \multicolumn{1}{c}{10\%}&     3.23 &     3.24 &     3.22 &     3.30 &     4.17 \\
        \multicolumn{1}{c}{20\%}&     3.33 &     3.32 &     3.32 &     3.43 &     4.34 \\
        \multicolumn{1}{c}{30\%}&     3.36 &     3.37 &     3.39 &     3.52 &     4.43 \\
        \multicolumn{1}{c}{40\%}&     3.37 &     3.39 &     3.43 &     3.55 &     4.47 \\
        \multicolumn{1}{c}{50\%}&     3.34 &     3.37 &     3.40 &     3.52 &     4.46 \\
        \bottomrule
    \end{tabularx}
    \label{tab:tab_life_cycle_gains_tot}
\end{table}

\begin{table}[!t]
    \justifying{\noindent \footnotesize}\par
    \centering
    \caption*{\textbf{Table \ref{tab:tab_life_cycle_gains_tot}: Life-Cycle Gains (Continued)}}
    \footnotesize
    \begin{tabularx}{\textwidth}{p{.15\linewidth}YYYYY}
        \toprule
        \multicolumn{6}{c}{Panel E: Wealth Gains at Death, Saving Rate\,=\,20\%} \\
        \midrule
        & \multicolumn{5}{c}{Down Payment} \\
        \cmidrule{2-6}
        \multicolumn{1}{c}{Threshold}
        & 10\% & 20\% & 30\% & 50\% & 100\% \\
        \midrule
        \multicolumn{1}{c}{10\%}&     9.22 &     7.91 &     6.66 &     4.66 &     2.41 \\
        \multicolumn{1}{c}{20\%}&     8.70 &     7.54 &     6.51 &     4.96 &     3.30 \\
        \multicolumn{1}{c}{30\%}&     8.33 &     7.38 &     6.57 &     5.35 &     4.05 \\
        \multicolumn{1}{c}{40\%}&     8.22 &     7.45 &     6.80 &     5.78 &     4.69 \\
        \multicolumn{1}{c}{50\%}&     8.29 &     7.65 &     7.07 &     6.19 &     5.22 \\
        \midrule
        \multicolumn{6}{c}{Panel F: CEVs at Death, Saving Rate\,=\,20\%} \\
        \midrule
        & \multicolumn{5}{c}{Down Payment} \\
        \cmidrule{2-6}
        \multicolumn{1}{c}{Threshold}
        & 10\% & 20\% & 30\% & 50\% & 100\% \\
        \midrule
        \multicolumn{1}{c}{10\%}&     3.97 &     4.24 &     4.47 &     4.90 &     6.23 \\
        \multicolumn{1}{c}{20\%}&     4.17 &     4.38 &     4.56 &     4.96 &     6.20 \\
        \multicolumn{1}{c}{30\%}&     4.24 &     4.41 &     4.58 &     4.95 &     6.14 \\
        \multicolumn{1}{c}{40\%}&     4.21 &     4.37 &     4.54 &     4.88 &     6.05 \\
        \multicolumn{1}{c}{50\%}&     4.12 &     4.28 &     4.43 &     4.76 &     5.94 \\
        \bottomrule
    \end{tabularx}
\end{table}

\subsection{Homeownership as a Life-Cycle Goldmine}
\label{subsec:life_cycle_gains_in_wealth_and_welfare}
\noindent Table~\ref{tab:tab_life_cycle_gains_tot} reports the life-cycle wealth gains and consumption-equivalent gains at death from single-home ownership. We measure the wealth gains as the percentage difference in mean wealth across the households in each strategy relative to the corresponding all-equity strategy:
\begin{align}
\label{eq: wealth_chg}
\Delta \bar{W}_{\tau,j}&=\frac{\bar{W}_{\tau,j}}{\bar{W}_{\tau,equity}}-1
\end{align}
 $\bar{W}_{\tau,j}$ is the mean dollar-value wealth of homeowner strategy $j$ at life-cycle time $\tau\in\{\text{retirement, death}\}$ and $\bar{W}_{\tau,equity}$ is the corresponding quantity for the all-equity strategy. 

We measure welfare gains using CEV, following \citet{cocco2005consumption}. Let $T_{i,\tau}$ denote household $i$'s
age at horizon $\tau$, with $T_{i,\text{death}}=T_i$, and define lifetime
utility over horizon $\tau$ as
\begin{align}
\label{eq: value}
V_{i,\tau}(\pi_j) = \sum_{t=0}^{T_{i,\tau}}\beta^t
\frac{c_{it}(\pi_j)^{1-\delta}}{1-\delta}
+ \mathbf{1}\{\tau=\text{death}\}\,\beta^{T_i} a_q
\frac{\bigl(W_{iT_i}(\pi_j)+b_q\bigr)^{1-\delta}}{1-\delta},
\end{align}
where $c_{it}(\pi_j)$ is realized adult-equivalent consumption at age $t$ under
strategy $\pi_j$, $W_{iT_i}(\pi_j)$ is wealth at death, and $a_q$ and $b_q$
are the bequest parameters calibrated in Section \ref{subsec:parameterization}. 
CEV is the constant consumption level that delivers the same utility over the horizon $\tau$:
\begin{align}
\label{eq: equiw}
\bar{c}_{i,\tau}(\pi_j)=\left[\frac{(1-\delta)\,V_{i,\tau}(\pi_j)}
{\sum_{t=0}^{T_{i,\tau}}\beta^t}\right]^{\frac{1}{1-\delta}}
\end{align}
The household-level CEV relative to the matched all-equity renter is
\begin{align}
\label{eq: welfare_chg}
\Delta c_{i,\tau,j}&=\frac{\bar{c}_{i,\tau}(\pi_j)}{\bar{c}_{i,\tau}(\pi^{equity}_i)}-1,
\end{align}
and the reported strategy-level consumption-equivalent gain is the cross-household average
\begin{align}
\Delta \bar{c}_{\tau,j}&=\frac{1}{N_j}\sum_{i=1}^{N_j}\Delta c_{i,\tau,j}.
\end{align}


Panels~A, C, and E of Table~\ref{tab:tab_life_cycle_gains_tot} document wealth gains across saving rates. Despite the historical underperformance of real estate relative to stocks, most homeowner strategies generate substantial terminal-wealth gains relative to saving-rate-matched renters; at the baseline 15\% saving rate with a 10\% down payment and 10\% purchase threshold, the gain is 12.5\%. Two patterns shape these gains. First, wealth gains decline monotonically with the saving rate. Households that save less accumulate less wealth, so the leverage embedded in the housing position constitutes a larger share of the portfolio. Second, at a given saving rate, low-down-payment strategies deliver the largest wealth gains. The purchase threshold interacts with the down payment rather than shifting gains uniformly, and it does so consistently across all three saving rates: Higher thresholds reduce gains at low down payments by delaying leveraged exposure but raise gains at high down payments by strengthening balance-sheet buffers at entry. 

Panels~B, D, and F of Table~\ref{tab:tab_life_cycle_gains_tot} report household-level CEV gains. Adequate saving is a precondition for homeownership to generate welfare gains. At a 10\% saving rate, higher-down-payment rules produce negative CEV, as meeting purchase requirements with limited liquid savings suppresses working-life consumption. At saving rates of 15\% and above, CEV gains are positive across all strategies and rise with the down payment, peaking under full cash purchase. Wealth gains and consumption-equivalent gains therefore respond in opposite directions to the saving rate: Low saving rates maximize wealth gains because housing leverage represents a larger share of a smaller portfolio, while high saving rates maximize CEV gains  because stronger financial buffers prevent mortgage obligations from crowding out consumption. 

\subsection{Life-Cycle Gains Versus Common Renter Benchmarks}
\label{subsec:homeownership_as_life_cycle_goldmine_life_cycle_gains_versus_common_renter_benchmarks}

\noindent In Section~\ref{subsec:life_cycle_gains_in_wealth_and_welfare}, we benchmark each homeowner strategy against a saving-rate-matched all-equity renter, a comparison that isolates the effect of housing under the highest-return liquid alternative. Since renters in practice follow a variety of investment strategies, we now broaden the comparison to three widely used alternatives.

The 60/40 rule allocates 60\% of financial wealth to stocks and 40\% to bonds. The 100-minus-age rule sets the equity share, in percent, to 100 minus the household's age. The TDF glide path keeps equity at 90\% through age 40 and then steps it down to 60\%, 50\%, and 30\% at ages 60, 65, and 72. For each rule, we consider two implementations of the stock sleeve: Panels~A--F split it one-third domestic and two-thirds international, while Panels~G--L replace it with domestic equity only. Table~\ref{tab:tab_benchmark_all_strategies_tot_reorg} reports the resulting comparisons over the full threshold-by-down-payment grid at the baseline 15\% saving rate.

Three patterns are clear. First, homeownership dominates every common renter benchmark across the full strategy grid, and the ranking of benchmarks is stable: Gains are smallest against the 60/40 rule, larger against the TDF glide path, and largest against the 100-minus-age rule. At the 30\% threshold and 10\% down payment, for example, the homeowner's consumption-equivalent gain is 6.25\% against the 60/40 rule, 7.01\% against the TDF glide path, and 9.36\% against the 100-minus-age rule, with corresponding terminal-wealth gains of 81.04\%, 114.69\%, and 161.81\%. The magnitudes reflect the equity share of each benchmark: Renters holding less equity accumulate less wealth, which magnifies the relative value of the leveraged housing position. Second, within each benchmark, consumption-equivalent gains rise with the down payment, while wealth gains are largest at low down payments. The leverage-welfare trade-off documented in the baseline table therefore survives under every practical renter strategy. Third, replacing the diversified stock sleeve with domestic equity only (Panels~G--L) yields slightly larger consumption-equivalent gains and nearly identical benchmark rankings, indicating that the results do not depend on international diversification. These results also hold at saving rates of 10\% and 20\%.

Comparing these results to the all-equity strategy in Table~\ref{tab:tab_life_cycle_gains_tot} reinforces two conclusions. The all-equity renter remains the most demanding benchmark: At the same 30\% threshold and 10\% down payment, the homeowner's consumption-equivalent gain against the all-equity renter is only 3.36\%, well below the 6\%--9\% range against the common benchmarks. The all-equity renter is thus the cleanest yardstick for isolating the marginal value of housing since it holds the saving rate fixed and assigns the renter the highest-return liquid alternative. At the same time, the fact that homeownership dominates by an even wider margin against the practical strategies that renters actually pursue strengthens the case that the life-cycle gains from homeownership are economically meaningful in realistic settings.

\begin{table}[!t]
    \justifying{\noindent \footnotesize This table reports common-benchmark comparisons from stationary block-bootstrap simulations of 1~million household life cycles at the baseline 15\% saving rate. The homeowner and the matched renter face identical labor-income, asset-return, inflation, and survival paths and follow the same saving rule after fixed costs. The renter rents for life and invests only in financial assets; the homeowner purchases once financial wealth covers the ``Down Payment'' plus an additional purchase ``Threshold,'' both as percentages of home value. Renter benchmarks are the 60/40 rule, the 100-minus-age rule, and a target-date-fund (TDF) glide path (see text). Panels~A--F use a stock sleeve split one-third domestic and two-thirds international; Panels~G--L use a domestic-only sleeve. Within each benchmark group, the first panel reports wealth gains at death and the second reports consumption-equivalent gains at death. All outcomes are expressed relative to the saving-rate-matched common renter benchmarks. Wealth gains are percentage differences in mean wealth at death. Consumption-equivalent variation (CEV) is the percentage difference between the homeowner and renter benchmark strategies in the constant consumption levels that replicate each strategy's realized lifetime utility, averaged across households. A positive CEV is a consumption-equivalent gain. All gains are in percent.\vspace{4pt}}\par
    \centering
    \caption{\textbf{Homeownership Gains Versus Common Renter Benchmarks}}
    \footnotesize
    \begin{tabularx}{\textwidth}{p{.15\linewidth}YYYYY}
        \toprule
        \multicolumn{6}{c}{Panel A: Diversified Sleeve, 60/40 Stock-Bond, Wealth Gains at Death} \\
        \midrule
        & \multicolumn{5}{c}{Down Payment} \\
        \cmidrule{2-6}
        \multicolumn{1}{c}{Threshold}
        & 10\% & 20\% & 30\% & 50\% & 100\% \\
        \midrule
        \multicolumn{1}{c}{10\%}& 83.75 & 80.77 & 78.13 & 74.16 & 70.31 \\
        \multicolumn{1}{c}{20\%}& 82.17 & 79.70 & 77.63 & 74.66 & 72.08 \\
        \multicolumn{1}{c}{30\%}& 81.04 & 79.15 & 77.60 & 75.38 & 73.51 \\
        \multicolumn{1}{c}{40\%}& 80.62 & 79.18 & 77.95 & 76.15 & 74.64 \\
        \multicolumn{1}{c}{50\%}& 80.66 & 79.46 & 78.42 & 76.81 & 75.56 \\
        \midrule
        \multicolumn{6}{c}{Panel B: Diversified Sleeve, 60/40 Stock-Bond, Consumption-Equivalent Gains at Death} \\
        \midrule
        & \multicolumn{5}{c}{Down Payment} \\
        \cmidrule{2-6}
        \multicolumn{1}{c}{Threshold}
        & 10\% & 20\% & 30\% & 50\% & 100\% \\
        \midrule
        \multicolumn{1}{c}{10\%}& 6.12 & 6.12 & 6.10 & 6.16 & 7.04 \\
        \multicolumn{1}{c}{20\%}& 6.22 & 6.20 & 6.20 & 6.30 & 7.21 \\
        \multicolumn{1}{c}{30\%}& 6.25 & 6.25 & 6.27 & 6.39 & 7.31 \\
        \multicolumn{1}{c}{40\%}& 6.25 & 6.27 & 6.30 & 6.42 & 7.36 \\
        \multicolumn{1}{c}{50\%}& 6.22 & 6.24 & 6.28 & 6.39 & 7.35 \\
        \midrule
        \multicolumn{6}{c}{Panel C: Diversified Sleeve, 100-Minus-Age, Wealth Gains at Death} \\
        \midrule
        & \multicolumn{5}{c}{Down Payment} \\
        \cmidrule{2-6}
        \multicolumn{1}{c}{Threshold}
        & 10\% & 20\% & 30\% & 50\% & 100\% \\
        \midrule
        \multicolumn{1}{c}{10\%}& 165.73 & 161.42 & 157.60 & 151.86 & 146.30 \\
        \multicolumn{1}{c}{20\%}& 163.45 & 159.87 & 156.89 & 152.59 & 148.85 \\
        \multicolumn{1}{c}{30\%}& 161.81 & 159.07 & 156.84 & 153.63 & 150.92 \\
        \multicolumn{1}{c}{40\%}& 161.21 & 159.12 & 157.35 & 154.74 & 152.55 \\
        \multicolumn{1}{c}{50\%}& 161.27 & 159.53 & 158.02 & 155.69 & 153.88 \\
        \bottomrule
    \end{tabularx}
    \label{tab:tab_benchmark_all_strategies_tot_reorg}
\end{table}

\FloatBarrier
\begin{table}[!ht]
    \centering
    \caption*{\textbf{Table~\ref{tab:tab_benchmark_all_strategies_tot_reorg}}: Homeownership Gains Versus Common Renter Benchmarks (Continued)}
    \footnotesize
    \begin{tabularx}{\textwidth}{p{.15\linewidth}YYYYY}
        \toprule
        \multicolumn{6}{c}{Panel D: Diversified Sleeve, 100-Minus-Age, Consumption-Equivalent Gains at Death} \\
        \midrule
        & \multicolumn{5}{c}{Down Payment} \\
        \cmidrule{2-6}
        \multicolumn{1}{c}{Threshold}
        & 10\% & 20\% & 30\% & 50\% & 100\% \\
        \midrule
        \multicolumn{1}{c}{10\%}& 9.23 & 9.21 & 9.19 & 9.24 & 10.15 \\
        \multicolumn{1}{c}{20\%}& 9.32 & 9.29 & 9.29 & 9.39 & 10.33 \\
        \multicolumn{1}{c}{30\%}& 9.36 & 9.35 & 9.37 & 9.49 & 10.45 \\
        \multicolumn{1}{c}{40\%}& 9.36 & 9.38 & 9.41 & 9.53 & 10.49 \\
        \multicolumn{1}{c}{50\%}& 9.33 & 9.35 & 9.39 & 9.50 & 10.49 \\
        \midrule
        \multicolumn{6}{c}{Panel E: Diversified Sleeve, TDF Glide Path, Wealth Gains at Death} \\
        \midrule
        & \multicolumn{5}{c}{Down Payment} \\
        \cmidrule{2-6}
        \multicolumn{1}{c}{Threshold}
        & 10\% & 20\% & 30\% & 50\% & 100\% \\
        \midrule
        \multicolumn{1}{c}{10\%}& 117.91 & 114.37 & 111.24 & 106.54 & 101.98 \\
        \multicolumn{1}{c}{20\%}& 116.04 & 113.11 & 110.66 & 107.13 & 104.07 \\
        \multicolumn{1}{c}{30\%}& 114.69 & 112.45 & 110.62 & 107.99 & 105.76 \\
        \multicolumn{1}{c}{40\%}& 114.20 & 112.49 & 111.04 & 108.90 & 107.10 \\
        \multicolumn{1}{c}{50\%}& 114.25 & 112.83 & 111.59 & 109.68 & 108.20 \\
        \midrule
        \multicolumn{6}{c}{Panel F: Diversified Sleeve, TDF Glide Path, Consumption-Equivalent Gains at Death} \\
        \midrule
        & \multicolumn{5}{c}{Down Payment} \\
        \cmidrule{2-6}
        \multicolumn{1}{c}{Threshold}
        & 10\% & 20\% & 30\% & 50\% & 100\% \\
        \midrule
        \multicolumn{1}{c}{10\%}& 6.88 & 6.87 & 6.85 & 6.91 & 7.80 \\
        \multicolumn{1}{c}{20\%}& 6.97 & 6.95 & 6.95 & 7.05 & 7.98 \\
        \multicolumn{1}{c}{30\%}& 7.01 & 7.00 & 7.03 & 7.15 & 8.09 \\
        \multicolumn{1}{c}{40\%}& 7.01 & 7.03 & 7.06 & 7.18 & 8.13 \\
        \multicolumn{1}{c}{50\%}& 6.98 & 7.01 & 7.04 & 7.15 & 8.12 \\
        \midrule
        \multicolumn{6}{c}{Panel G: Domestic Equity Sleeve, 60/40 Stock-Bond, Wealth Gains at Death} \\
        \midrule
        & \multicolumn{5}{c}{Down Payment} \\
        \cmidrule{2-6}
        \multicolumn{1}{c}{Threshold}
        & 10\% & 20\% & 30\% & 50\% & 100\% \\
        \midrule
        \multicolumn{1}{c}{10\%}& 84.08 & 81.09 & 78.44 & 74.47 & 70.62 \\
        \multicolumn{1}{c}{20\%}& 82.50 & 80.02 & 77.95 & 74.97 & 72.38 \\
        \multicolumn{1}{c}{30\%}& 81.36 & 79.47 & 77.92 & 75.70 & 73.82 \\
        \multicolumn{1}{c}{40\%}& 80.94 & 79.50 & 78.27 & 76.47 & 74.95 \\
        \multicolumn{1}{c}{50\%}& 80.99 & 79.78 & 78.74 & 77.12 & 75.87 \\
        \midrule
        \multicolumn{6}{c}{Panel H: Domestic Equity Sleeve, 60/40 Stock-Bond, Consumption-Equivalent Gains at Death} \\
        \midrule
        & \multicolumn{5}{c}{Down Payment} \\
        \cmidrule{2-6}
        \multicolumn{1}{c}{Threshold}
        & 10\% & 20\% & 30\% & 50\% & 100\% \\
        \midrule
        \multicolumn{1}{c}{10\%}& 6.86 & 6.85 & 6.83 & 6.89 & 7.78 \\
        \multicolumn{1}{c}{20\%}& 6.95 & 6.93 & 6.93 & 7.03 & 7.95 \\
        \multicolumn{1}{c}{30\%}& 6.98 & 6.98 & 7.00 & 7.12 & 8.06 \\
        \multicolumn{1}{c}{40\%}& 6.99 & 7.01 & 7.04 & 7.16 & 8.10 \\
        \multicolumn{1}{c}{50\%}& 6.95 & 6.98 & 7.01 & 7.12 & 8.09 \\
        \bottomrule
    \end{tabularx}
\end{table}
\FloatBarrier

\FloatBarrier
\begin{table}[!ht]
    \centering
    \caption*{\textbf{Table~\ref{tab:tab_benchmark_all_strategies_tot_reorg}}: Homeownership Gains Versus Common Renter Benchmarks (Continued)}
    \footnotesize
    \begin{tabularx}{\textwidth}{p{.15\linewidth}YYYYY}
        \toprule
        \multicolumn{6}{c}{Panel I: Domestic Equity Sleeve, 100-Minus-Age, Wealth Gains at Death} \\
        \midrule
        & \multicolumn{5}{c}{Down Payment} \\
        \cmidrule{2-6}
        \multicolumn{1}{c}{Threshold}
        & 10\% & 20\% & 30\% & 50\% & 100\% \\
        \midrule
        \multicolumn{1}{c}{10\%}& 162.89 & 158.62 & 154.84 & 149.17 & 143.66 \\
        \multicolumn{1}{c}{20\%}& 160.63 & 157.09 & 154.14 & 149.89 & 146.18 \\
        \multicolumn{1}{c}{30\%}& 159.01 & 156.30 & 154.09 & 150.92 & 148.23 \\
        \multicolumn{1}{c}{40\%}& 158.41 & 156.35 & 154.60 & 152.02 & 149.85 \\
        \multicolumn{1}{c}{50\%}& 158.47 & 156.75 & 155.26 & 152.96 & 151.17 \\
        \midrule
        \multicolumn{6}{c}{Panel J: Domestic Equity Sleeve, 100-Minus-Age, Consumption-Equivalent Gains at Death} \\
        \midrule
        & \multicolumn{5}{c}{Down Payment} \\
        \cmidrule{2-6}
        \multicolumn{1}{c}{Threshold}
        & 10\% & 20\% & 30\% & 50\% & 100\% \\
        \midrule
        \multicolumn{1}{c}{10\%}& 9.55 & 9.54 & 9.51 & 9.57 & 10.47 \\
        \multicolumn{1}{c}{20\%}& 9.65 & 9.62 & 9.62 & 9.72 & 10.66 \\
        \multicolumn{1}{c}{30\%}& 9.68 & 9.68 & 9.70 & 9.82 & 10.77 \\
        \multicolumn{1}{c}{40\%}& 9.69 & 9.71 & 9.74 & 9.86 & 10.82 \\
        \multicolumn{1}{c}{50\%}& 9.66 & 9.68 & 9.71 & 9.83 & 10.82 \\
        \midrule
        \multicolumn{6}{c}{Panel K: Domestic Equity Sleeve, TDF Glide Path, Wealth Gains at Death} \\
        \midrule
        & \multicolumn{5}{c}{Down Payment} \\
        \cmidrule{2-6}
        \multicolumn{1}{c}{Threshold}
        & 10\% & 20\% & 30\% & 50\% & 100\% \\
        \midrule
        \multicolumn{1}{c}{10\%}& 116.07 & 112.56 & 109.45 & 104.79 & 100.27 \\
        \multicolumn{1}{c}{20\%}& 114.21 & 111.30 & 108.87 & 105.38 & 102.34 \\
        \multicolumn{1}{c}{30\%}& 112.87 & 110.65 & 108.84 & 106.23 & 104.02 \\
        \multicolumn{1}{c}{40\%}& 112.39 & 110.69 & 109.25 & 107.13 & 105.35 \\
        \multicolumn{1}{c}{50\%}& 112.44 & 111.02 & 109.80 & 107.90 & 106.43 \\
        \midrule
        \multicolumn{6}{c}{Panel L: Domestic Equity Sleeve, TDF Glide Path, Consumption-Equivalent Gains at Death} \\
        \midrule
        & \multicolumn{5}{c}{Down Payment} \\
        \cmidrule{2-6}
        \multicolumn{1}{c}{Threshold}
        & 10\% & 20\% & 30\% & 50\% & 100\% \\
        \midrule
        \multicolumn{1}{c}{10\%}& 7.50 & 7.49 & 7.46 & 7.52 & 8.42 \\
        \multicolumn{1}{c}{20\%}& 7.59 & 7.57 & 7.57 & 7.67 & 8.60 \\
        \multicolumn{1}{c}{30\%}& 7.62 & 7.62 & 7.64 & 7.76 & 8.71 \\
        \multicolumn{1}{c}{40\%}& 7.63 & 7.65 & 7.68 & 7.80 & 8.75 \\
        \multicolumn{1}{c}{50\%}& 7.60 & 7.62 & 7.66 & 7.77 & 8.74 \\
        \bottomrule    
    \end{tabularx}
\end{table}
\FloatBarrier


\FloatBarrier
\begin{table}[!ht]
    \justifying{\noindent \footnotesize This table reports welfare-decomposition diagnostics from stationary block-bootstrap simulations of 1~million household life cycles at the baseline 15\% saving rate. The homeowner and the matched renter face identical labor-income, asset-return, inflation, and survival paths and follow the same saving rule after fixed costs. The renter rents for life and invests only in stocks; the homeowner purchases once financial wealth covers the ``Down Payment'' plus an additional purchase ``Threshold,'' both as percentages of home value. Panel~A reports consumption-equivalent gains calculated based on the consumption component of lifetime utility. Panel~B reports the analogous consumption-equivalent gains from the bequest component only. All outcomes are expressed relative to the saving-rate-matched all-equity renter benchmark. Consumption-equivalent variation (CEV) is the percentage difference between the homeowner and renter benchmark strategies in the constant consumption levels that replicate each strategy's realized component utility, i.e., consumption component or bequest component, averaged across households. A positive CEV is a consumption-equivalent gain. All gains are in percent.\vspace{4pt}}\par
    \centering
    \caption{\textbf{Welfare Decomposition}}
    \footnotesize
    \begin{tabularx}{\textwidth}{p{.15\linewidth}YYYYY}
        \toprule
        \multicolumn{6}{c}{Panel A: Consumption Channel, Consumption-Equivalent Gains ($\bar{c}^c$)} \\
\midrule
& \multicolumn{5}{c}{Down Payment} \\
\cmidrule{2-6}
\multicolumn{1}{c}{Threshold}
& 10\% & 20\% & 30\% & 50\% & 100\% \\
\midrule
\multicolumn{1}{c}{10\%}&     2.19 &     2.30 &     2.34 &     2.46 &     3.28 \\
\multicolumn{1}{c}{20\%}&     2.42 &     2.46 &     2.50 &     2.64 &     3.49 \\
\multicolumn{1}{c}{30\%}&     2.56 &     2.59 &     2.64 &     2.80 &     3.64 \\
\multicolumn{1}{c}{40\%}&     2.65 &     2.70 &     2.75 &     2.90 &     3.74 \\
\multicolumn{1}{c}{50\%}&     2.69 &     2.74 &     2.80 &     2.94 &     3.79 \\
\midrule
\multicolumn{6}{c}{Panel B: Bequest Channel, Consumption-Equivalent Gains ($\bar{c}^W$)} \\
\midrule
& \multicolumn{5}{c}{Down Payment} \\
\cmidrule{2-6}
\multicolumn{1}{c}{Threshold}
& 10\% & 20\% & 30\% & 50\% & 100\% \\
\midrule
\multicolumn{1}{c}{10\%}&    11.39 &     9.76 &     8.32 &     6.16 &     4.06 \\
\multicolumn{1}{c}{20\%}&    10.53 &     9.17 &     8.05 &     6.43 &     5.02 \\
\multicolumn{1}{c}{30\%}&     9.90 &     8.87 &     8.03 &     6.82 &     5.81 \\
\multicolumn{1}{c}{40\%}&     9.67 &     8.89 &     8.22 &     7.24 &     6.42 \\
\multicolumn{1}{c}{50\%}&     9.69 &     9.03 &     8.47 &     7.59 &     6.92 \\
        \bottomrule
    \end{tabularx}
    \label{tab:tab_dissect_welfare_tot_sr15}
\end{table}
\FloatBarrier

\justifying
\subsection{Welfare Decomposition}
\label{subsec:welfare_decomposition}
\justifying \noindent While wealth gains reflect differences in asset accumulation, welfare depends on how resources translate into consumption and bequests over the life cycle. Table~\ref{tab:tab_dissect_welfare_tot_sr15} therefore decomposes the welfare gain at a 15\% saving rate into a consumption channel and a bequest channel. Panel~A reports the percentage gain in $\bar{c}^c$, the consumption equivalent implied by the consumption component $V_i^c(\pi_j)$ alone, and Panel~B reports the analogous gain in $\bar{c}^W$, implied by the bequest component $V_i^W(\pi_j)$. Both are obtained from the structural inversion in Equation~\ref{eq: equiw}.\footnote{The two channel-level consumption equivalents are informative about the sources of the aggregate welfare gain but are not additive components of it. We report them separately because the standard additive aggregation is sensitive to the discount factor and bequest parameters, which can mechanically attenuate the bequest channel and cause the consumption channel to dominate the overall value function.}

Homeownership improves welfare through both channels. The consumption equivalent $\bar{c}^c$ rises by 2\% to 4\% across all strategy cells: Although homeownership depresses consumption during the mortgage-service years, it improves the intertemporal allocation of consumption overall. This result is inconsistent with the view that homeownership lowers retirement consumption by immobilizing resources in an illiquid asset, and it reflects the liquidation rules and the reverse mortgage option, both of which release housing wealth later in life. The bequest channel is larger: $\bar{c}^W$ rises by 4\% to 11\%, with the largest gains at higher leverage, as accumulated home equity is transferred to heirs. Together, these results reconcile the wealth and welfare findings: Homeownership can reduce liquid financial wealth and working-life consumption while still raising the total consumption-equivalent gain  because it delivers higher retirement consumption and larger bequests.

\begin{table}[!ht]
    \justifying{\noindent \footnotesize This table reports heterogeneous outcomes from stationary block-bootstrap simulations of 1~million household life cycles at the baseline 15\% saving rate, averaged across the full 25-cell threshold-by-down-payment grid. The homeowner and the matched renter face identical labor-income, asset-return, inflation, and survival paths and follow the same saving rule after fixed costs. The renter rents for life and invests only in stocks; the homeowner purchases once financial wealth covers the down payment plus an additional purchase threshold, both as percentages of home value. Within each homeowner strategy, households are sorted into quintiles by labor income (Panel~A), house price at purchase (Panel~B), and mortgage rate at purchase (Panel~C), and the resulting quintile statistics are averaged across strategies. Q1 denotes the lowest quintile and Q5 the highest. All outcomes are expressed relative to the saving-rate-matched all-equity renter benchmark. ``Wealth Loss at Retirement'' and ``Wealth Gain at Death'' are the percentage differences in mean net worth at retirement and at death, respectively. ``Consumption-Equivalent Gain at Death'' reports consumption-equivalent gains at death. Consumption-equivalent variation (CEV) is the percentage difference between the homeowner and renter benchmark strategies in the constant consumption levels that replicate each strategy's realized lifetime utility, averaged across households. A positive CEV is a consumption-equivalent gain. All values are in percent.\vspace{4pt}}\par
    \centering
    \caption{\textbf{Heterogeneity}}
    \footnotesize
    \begin{tabularx}{\textwidth}{YYYY}
        \toprule
        \multicolumn{4}{c}{Panel A: Labor Income Quintiles} \\
        \midrule
        & Wealth Loss at Retirement & Wealth Gain at Death & Consumption-Equivalent Gain at Death \\
        \midrule
        \multicolumn{1}{c}{Q1} & -9.66 & 18.46 & 0.93 \\
        \multicolumn{1}{c}{Q2} & -11.44 & 14.82 & 2.93 \\
        \multicolumn{1}{c}{Q3} & -11.82 & 12.27 & 4.41 \\
        \multicolumn{1}{c}{Q4} & -11.75 & 9.53 & 5.72 \\
        \multicolumn{1}{c}{Q5} & -10.22 & 5.05 & 6.57 \\
        \midrule
        \multicolumn{4}{c}{Panel B: Housing Price Quintiles} \\
        \midrule
        & Wealth Loss at Retirement & Wealth Gain at Death & Consumption-Equivalent Gain at Death \\
        \midrule
        \multicolumn{1}{c}{Q1} & -8.95 & 3.30 & 3.62 \\
        \multicolumn{1}{c}{Q2} & -11.09 & 5.72 & 4.07 \\
        \multicolumn{1}{c}{Q3} & -12.28 & 8.48 & 4.21 \\
        \multicolumn{1}{c}{Q4} & -12.81 & 12.81 & 4.11 \\
        \multicolumn{1}{c}{Q5} & -10.54 & 24.97 & 4.56 \\
        \midrule
        \multicolumn{4}{c}{Panel C: Mortgage-Rate Quintiles} \\
        \midrule
        & Wealth Loss at Retirement & Wealth Gain at Death & Consumption-Equivalent Gain at Death \\
        \midrule
        \multicolumn{1}{c}{Q1} & -5.92 & 16.37 & 5.35 \\
        \multicolumn{1}{c}{Q2} & -11.03 & 8.16 & 3.69 \\
        \multicolumn{1}{c}{Q3} & -13.39 & 6.93 & 3.77 \\
        \multicolumn{1}{c}{Q4} & -9.75 & 11.22 & 4.21 \\
        \multicolumn{1}{c}{Q5} & -13.54 & 4.24 & 3.57 \\
        \bottomrule
    \end{tabularx}
    \label{tab:tab_hetero_avg_sr15}
\end{table}

\justifying
\subsection{Household Heterogeneity}
\label{subsec:household_heterogeneity}
\noindent Table~\ref{tab:tab_hetero_avg_sr15} reports heterogeneous effects of homeownership, averaged  across all 25 owner strategies at the 15\% saving rate. Within each strategy cell, we sort households into quintiles by labor income, house price at purchase, and mortgage rate at purchase,   then average the quintile statistics across the full threshold-by-down-payment grid. For each quintile, the table reports the wealth change at retirement, the wealth change at death, and the lifetime consumption-equivalent gain .

Panel~A reveals an income reversal between the wealth and welfare margins. Wealth gains at death decline monotonically with labor income, from 18.46\% in the lowest quintile to 5.05\% in the highest, whereas lifetime consumption-equivalent gains rise from 0.93\% to 6.57\%. The reversal is not driven by differential accumulation before retirement: All income groups give up wealth at retirement relative to their matched renters. The pattern instead reflects two offsetting forces. Housing occupies a larger share of low-income households' balance sheets, so housing leverage contributes more to their late-life wealth accumulation; but the same households have less capacity to absorb mortgage-service pressure during working life, which compresses their consumption gains. For high-income households, housing is a smaller component of a larger lifetime portfolio, so terminal-wealth gains are modest; but stronger income flows support a smoother consumption path through the mortgage years and into retirement, which raises the consumption-equivalent gain.


Panel~B shows that the house-price environment at purchase also generates substantial dispersion in homeowner gains. Households purchasing in the highest house-price quintile record the largest end-of-life gains on both margins, with a wealth gain at death of 24.97\% and a consumption-equivalent gain of 4.56\%, compared with 3.30\% and 3.62\% in the lowest house-price quintile. This ranking is absent at retirement, where the highest-price quintile fares slightly worse than the lowest; the advantage of purchasing in high-price environments emerges only later in life, as stronger housing-linked wealth accumulation outweighs a weaker retirement position. One interpretation is that high-house-price environments tend to coincide with stronger macroeconomic conditions and more favorable long-run income realizations, which allow  households to sustain mortgage obligations while home equity compounds on a higher price base.


Panel~C shows that financing conditions at purchase remain quantitatively important after averaging across strategies. Households entering in the lowest mortgage-rate quintile gain 16.37\% in wealth at death and 5.35\% in lifetime consumption-equivalent terms, compared with 4.24\% and 3.57\% in the highest-rate quintile. The retirement comparison is sharper still: The  lowest-rate group forgoes only 5.92\% of retirement wealth relative to its matched renter, whereas the highest-rate group forgoes 13.54\%. Low-rate entry thus substantially reduces the working-life liquidity cost of ownership and preserves more balance-sheet capacity before retirement. 

Taken together, the results show that homeowner gains depend systematically on household income, purchase-time house prices, and purchase-time financing conditions. Across all three sorts, dispersion in the consumption-equivalent gain is considerably narrower than dispersion in terminal wealth, underscoring that the wealth and welfare margins need not move together across these environments.



\justifying
\subsection{Regional Data Evidence}
\label{subsec:regional_evidence}
\noindent We replicate the analysis separately for the U.S., the U.K., and continental Europe using the corresponding geographic subsamples of the Macrohistory Database.\footnote{For the continental European simulations, we use a hierarchical sampling scheme in which we first select a country uniformly at random from the continental European subsample and then draw a block of consecutive observations from that country, preserving within-country autocorrelation while incorporating cross-country variation across draws.} Household decision rules are held fixed at their U.S.-calibrated values across all regional simulations. The regional results serve as a robustness check; the main analysis uses global data both because it offers richer variation and because it mitigates lucky-market bias.


Panels~A, C, and E of Table~\ref{tab:tab_life_cycle_gains_regional_sr15} report mean wealth gains at death at the 15\% saving rate. The regional patterns closely mirror the global benchmark. Mortgage strategies generate wealth gains between roughly 6\% and 13\% across all three regions, while cash-purchase strategies range from 4\% to 8\%. The cross-regional spread within any given strategy cell is small: Continental Europe lies at most less than 1\% below the U.S. and U.K., which are essentially indistinguishable from each other. In every region, wealth gains are largest for leveraged strategies with low down payments and decline as the household substitutes equity for leverage. The wealth-accumulation advantage of homeownership is therefore not driven by any single national housing market.


Panels~B, D, and F of Table~\ref{tab:tab_life_cycle_gains_regional_sr15} report the average household-level CEV at the 15\% saving rate. All three regions again reproduce the global conclusions. CEV values are positive across the full strategy grid in every region, and the monotonic relationship between the down payment and  CEV is broadly preserved throughout. Mortgage-strategy gains cluster near 3.1\%--3.6\% and cash-purchase gains near 4.1\%--4.6\% in all three regions. The U.K. and continental Europe fall within 0.2\% of the global result across every strategy cell. The U.S. tracks the global result most closely, consistent with its dominant weight in the global sample. The welfare case for homeownership is therefore not specific to any particular national housing market.


\begin{table}[!ht]
    \justifying{\noindent \footnotesize This table reports regional outcomes from stationary block-bootstrap simulations of 1~million household life cycles at the baseline 15\% saving rate. The homeowner and the matched renter face identical labor-income, asset-return, inflation, and survival paths and follow the same saving rule after fixed costs. The renter rents for life and invests only in stocks; the homeowner purchases once financial wealth covers the ``Down Payment'' plus an additional purchase ``Threshold,'' both as percentages of home value. Each regional block uses the corresponding geographic subsample of the Macrohistory Database \citep{jorda2019rate} while holding U.S. household decision rules fixed. Panels~A and B report wealth gains at death and average household-level consumption-equivalent gains at death using U.S. data; Panels~C and D report the corresponding outcomes using U.K. data; and Panels~E and F report the corresponding outcomes using continental European data. All outcomes are expressed relative to the saving-rate-matched all-equity renter benchmark. Wealth gains are percentage differences in mean wealth at death. Consumption-equivalent variation (CEV) is the percentage difference between the homeowner and renter benchmark strategies in the constant consumption levels that replicate each strategy's realized lifetime utility, averaged across households. A positive CEV is a consumption-equivalent gain. All gains are in percent.\vspace{4pt}}\par
    \caption{\textbf{Life-Cycle Gains with Regional Data}}
    \centering
    \footnotesize
    \begin{tabularx}{\textwidth}{p{.15\linewidth}YYYYY}
        \toprule
\multicolumn{6}{c}{Panel A: U.S. Wealth Gains at Death} \\
\midrule
& \multicolumn{5}{c}{Down Payment} \\
\cmidrule{2-6}
\multicolumn{1}{c}{Threshold}
& 10\% & 20\% & 30\% & 50\% & 100\% \\
\midrule
\multicolumn{1}{c}{10\%}&    12.52 &    10.75 &     9.18 &     6.83 &     4.81 \\
\multicolumn{1}{c}{20\%}&    11.56 &    10.10 &     8.88 &     7.17 &     5.94 \\
\multicolumn{1}{c}{30\%}&    10.88 &     9.79 &     8.91 &     7.65 &     6.85 \\
\multicolumn{1}{c}{40\%}&    10.68 &     9.85 &     9.17 &     8.15 &     7.56 \\
\multicolumn{1}{c}{50\%}&    10.74 &    10.06 &     9.46 &     8.56 &     8.12 \\
\midrule
\multicolumn{6}{c}{Panel B: U.S. Consumption-Equivalent Gains at Death} \\
\midrule
& \multicolumn{5}{c}{Down Payment} \\
\cmidrule{2-6}
\multicolumn{1}{c}{Threshold}
& 10\% & 20\% & 30\% & 50\% & 100\% \\
\midrule
\multicolumn{1}{c}{10\%}&     3.27 &     3.27 &     3.28 &     3.38 &     4.32 \\
\multicolumn{1}{c}{20\%}&     3.36 &     3.36 &     3.38 &     3.51 &     4.48 \\
\multicolumn{1}{c}{30\%}&     3.39 &     3.40 &     3.45 &     3.59 &     4.57 \\
\multicolumn{1}{c}{40\%}&     3.39 &     3.42 &     3.47 &     3.61 &     4.59 \\
\multicolumn{1}{c}{50\%}&     3.35 &     3.38 &     3.44 &     3.57 &     4.57 \\
\midrule
\multicolumn{6}{c}{Panel C: U.K. Wealth Gains at Death} \\
\midrule
& \multicolumn{5}{c}{Down Payment} \\
\cmidrule{2-6}
\multicolumn{1}{c}{Threshold}
& 10\% & 20\% & 30\% & 50\% & 100\% \\
\midrule
\multicolumn{1}{c}{10\%}&    12.79 &    10.96 &     9.31 &     6.78 &     4.34 \\
\multicolumn{1}{c}{20\%}&    11.81 &    10.27 &     8.98 &     7.08 &     5.46 \\
\multicolumn{1}{c}{30\%}&    11.11 &     9.91 &     8.94 &     7.51 &     6.37 \\
\multicolumn{1}{c}{40\%}&    10.83 &     9.91 &     9.13 &     7.97 &     7.09 \\
\multicolumn{1}{c}{50\%}&    10.82 &    10.06 &     9.38 &     8.38 &     7.67 \\
        \bottomrule
    \end{tabularx}
    \label{tab:tab_life_cycle_gains_regional_sr15}
\end{table}

\begin{table}[!ht]
    \justifying{\noindent \footnotesize}\par
    \centering
    \caption*{\textbf{Table \ref{tab:tab_life_cycle_gains_regional_sr15}: Life-Cycle Gains with Regional Data (Continued)}}
    \footnotesize
    \begin{tabularx}{\textwidth}{p{.15\linewidth}YYYYY}
        \toprule
\multicolumn{6}{c}{Panel D: U.K. Consumption-Equivalent Gains at Death} \\
\midrule
& \multicolumn{5}{c}{Down Payment} \\
\cmidrule{2-6}
\multicolumn{1}{c}{Threshold}
& 10\% & 20\% & 30\% & 50\% & 100\% \\
\midrule
\multicolumn{1}{c}{10\%}&     3.14 &     3.15 &     3.14 &     3.21 &     4.12 \\
\multicolumn{1}{c}{20\%}&     3.23 &     3.22 &     3.24 &     3.34 &     4.29 \\
\multicolumn{1}{c}{30\%}&     3.25 &     3.26 &     3.29 &     3.41 &     4.39 \\
\multicolumn{1}{c}{40\%}&     3.25 &     3.27 &     3.31 &     3.42 &     4.43 \\
\multicolumn{1}{c}{50\%}&     3.20 &     3.22 &     3.26 &     3.38 &     4.42 \\
\midrule
\multicolumn{6}{c}{Panel E: Continental Europe Wealth Gains at Death} \\
\midrule
& \multicolumn{5}{c}{Down Payment} \\
\cmidrule{2-6}
\multicolumn{1}{c}{Threshold}
& 10\% & 20\% & 30\% & 50\% & 100\% \\
\midrule
\multicolumn{1}{c}{10\%}&    12.10 &    10.28 &     8.66 &     6.26 &     3.94 \\
\multicolumn{1}{c}{20\%}&    11.14 &     9.63 &     8.38 &     6.58 &     5.03 \\
\multicolumn{1}{c}{30\%}&    10.46 &     9.30 &     8.37 &     7.04 &     5.90 \\
\multicolumn{1}{c}{40\%}&    10.22 &     9.33 &     8.60 &     7.52 &     6.61 \\
\multicolumn{1}{c}{50\%}&    10.25 &     9.51 &     8.90 &     7.95 &     7.17 \\
\midrule
\multicolumn{6}{c}{Panel F: Continental Europe Consumption-Equivalent Gains at Death} \\
\midrule
& \multicolumn{5}{c}{Down Payment} \\
\cmidrule{2-6}
\multicolumn{1}{c}{Threshold}
& 10\% & 20\% & 30\% & 50\% & 100\% \\
\midrule
\multicolumn{1}{c}{10\%}&     3.23 &     3.22 &     3.21 &     3.29 &     4.18 \\
\multicolumn{1}{c}{20\%}&     3.33 &     3.30 &     3.32 &     3.44 &     4.34 \\
\multicolumn{1}{c}{30\%}&     3.35 &     3.36 &     3.40 &     3.53 &     4.44 \\
\multicolumn{1}{c}{40\%}&     3.36 &     3.39 &     3.44 &     3.56 &     4.48 \\
\multicolumn{1}{c}{50\%}&     3.33 &     3.36 &     3.41 &     3.53 &     4.47 \\
\bottomrule
\end{tabularx}
\end{table}

\justifying
\subsection{Downside Risks}
\label{subsec:downside_risks}
\noindent Homeownership reduces portfolio volatility by replacing a fraction of equity with an asset that is less volatile and whose returns are less correlated with equity returns, and this benefit survives mortgage leverage. It is reflected in two downside-risk outcomes, reported in  Table~\ref{tab:tab_wealth_default_tot_sr15} relative to the all-equity renter benchmark. Panel~A reports the improvement in maximum portfolio draw-down. Homeownership reduces peak-to-trough wealth losses by 20.6\% to 24.4\% across the strategy grid at the 15\% saving rate, with the largest reductions for higher-equity, lower-threshold strategies. Panel~B reports the financial-ruin rate, defined as the fraction of households unable to afford the minimum consumption level at least once during the life cycle. Homeownership lowers this rate by the equivalent of 10 to 13 households out of 100 relative to the all-equity renter. The relevant consideration for households is therefore not the mean return alone but the joint distribution of returns and cash-flow needs over the life cycle. Housing performs well on this dimension because it trims the left tail of outcomes at the moments when households are most vulnerable, which helps explain why ownership can dominate on a welfare basis even though housing underperforms stocks in raw long-run return comparisons.


\begin{table}[ht]
    \justifying{\noindent \footnotesize This table reports downside-risk outcomes from stationary block-bootstrap simulations of 1~million household life cycles at the baseline 15\% saving rate. The homeowner and the matched renter face identical labor-income, asset-return, inflation, and survival paths and follow the same saving rule after fixed costs. The renter rents for life and invests only in stocks; the homeowner purchases once financial wealth covers the ``Down Payment'' plus an additional purchase ``Threshold,'' both as percentages of home value. Panel~A reports maximum-drawdown improvements of the homeowner strategy relative to the matched all-equity renter benchmark, defined as percentage differences in maximum drawdown, which is the largest peak-to-trough decline in the household's wealth over the life cycle. Panel~B reports financial-ruin reductions, where financial ruin occurs when a household cannot afford minimum consumption at least once during the life cycle. Each Panel~B entry reports the level difference in the financial-ruin rate between the homeowner and the matched all-equity renter benchmark, so an entry of $-12$ indicates that the homeowner financial-ruin rate is 0.12 lower than the renter rate. The all-equity renter benchmark's financial-ruin rate at the 15\% saving rate is 24.76\%. All values are in percent.\vspace{4pt}}\par
    \centering
    \caption{\textbf{Downside Risk Outcomes}}
    \footnotesize
    \begin{tabularx}{\textwidth}{p{.15\linewidth}YYYYY}
    \toprule
    \multicolumn{6}{c}{Panel A: Maximum Drawdown Improvement} \\
    \midrule
    & \multicolumn{5}{c}{Down Payment} \\
    \cmidrule{2-6}
    \multicolumn{1}{c}{Threshold}
    & 10\% & 20\% & 30\% & 50\% & 100\% \\
    \midrule
    \multicolumn{1}{c}{10\%}&    22.97 &    23.02 &    23.34 &    24.14 &    24.44 \\
    \multicolumn{1}{c}{20\%}&    22.58 &    22.71 &    23.03 &    23.58 &    23.88 \\
    \multicolumn{1}{c}{30\%}&    22.01 &    22.22 &    22.50 &    22.91 &    23.19 \\
    \multicolumn{1}{c}{40\%}&    21.37 &    21.59 &    21.81 &    22.10 &    22.43 \\
    \multicolumn{1}{c}{50\%}&    20.62 &    20.80 &    20.98 &    21.18 &    21.66 \\
    
    \midrule
    \multicolumn{6}{c}{Panel B: Financial-Ruin Reduction} \\
    \midrule
    & \multicolumn{5}{c}{Down Payment} \\
    \cmidrule{2-6}
    \multicolumn{1}{c}{Threshold}
    & 10\% & 20\% & 30\% & 50\% & 100\% \\
    \midrule
    \multicolumn{1}{c}{10\%}&   -12.48 &   -12.42 &   -12.30 &   -12.15 &   -12.56 \\
    \multicolumn{1}{c}{20\%}&   -11.94 &   -11.90 &   -11.85 &   -11.81 &   -12.36 \\
    \multicolumn{1}{c}{30\%}&   -11.41 &   -11.41 &   -11.39 &   -11.42 &   -12.01 \\
    \multicolumn{1}{c}{40\%}&   -10.93 &   -10.93 &   -10.91 &   -10.95 &   -11.58 \\
    \multicolumn{1}{c}{50\%}&   -10.40 &   -10.40 &   -10.39 &   -10.40 &   -11.12 \\
    
%
%
    \bottomrule
    \end{tabularx}
    \label{tab:tab_wealth_default_tot_sr15}
\end{table}

\justifying
\subsection{Wealth Inequality}
\label{subsec:wealth_inequality}
\justifying \noindent Since homeownership reduces portfolio volatility, it naturally compresses the cross-sectional distribution of wealth. We examine this effect through the Gini coefficient of household wealth at retirement. For strategy economy $j$, the Gini coefficient is
\begin{align}
G_j &= \frac{\sum_{m_i=1}^{n}\sum_{m_j=1}^{n} \abs{y_{m_i} - y_{m_j}}}{2n \sum_{m_i=1}^{n} y_{m_j}},
\end{align}
where $n = 1{,}000{,}000$ and $y_{m_j}$ is the retirement wealth of household $m_j$ in strategy $j$. Table~\ref{tab:tab_gini_tot_sr15} reports 100 times the Gini difference relative to the all-equity renter economy.

All homeowner strategies reduce wealth inequality at retirement. The homeowner Gini coefficient falls by up to 0.03 relative to the renter benchmark, a relative decline of roughly 7\%--8\%. Leveraged strategies deliver larger reductions than cash purchase, for two reinforcing reasons: The saving required before a cash purchase is attained sooner and more often by higher-income households, so the benefits of cash entry accrue more narrowly, while mortgage financing extends the wealth benefits of housing, amplified by leverage precisely for households with little initial wealth, to a broader set of households.  Expanding access to conventional-leverage mortgages therefore   materially lowers retirement-wealth inequality, consistent with the view that housing markets carry meaningful distributional consequences \citep{wolff70s_inequality,bhamra2019household}.


\FloatBarrier
\begin{table}[htbp!]
    \justifying{\footnotesize\noindent This table reports inequality outcomes from stationary block-bootstrap simulations of 1~million household life cycles at the baseline 15\% saving rate. The homeowner and the matched renter face identical labor-income, asset-return, inflation, and survival paths and follow the same saving rule after fixed costs. The renter rents for life and invests only in stocks; the homeowner purchases once financial wealth covers the ``Down Payment'' plus an additional purchase ``Threshold,'' both as percentages of home value. Each entry reports the level difference between the homeowner and the matched all-equity renter benchmark in the Gini coefficient for the retirement net-worth distribution, so an entry of $-3$ indicates that the homeowner Gini coefficient is 0.03 lower than the all-equity renter benchmark's. The all-equity renter benchmark's Gini coefficient at retirement is 0.4144. Entries are the Gini difference multiplied by 100.\vspace{4pt}}\par
    \centering
    \caption{\textbf{Gini Change at Retirement}}
    \footnotesize
    \begin{tabularx}{\textwidth}{p{.15\linewidth}YYYYY}
        \toprule
& \multicolumn{5}{c}{Down Payment} \\
\cmidrule{2-6}
\multicolumn{1}{c}{Threshold}
& 10\% & 20\% & 30\% & 50\% & 100\% \\
\midrule
\multicolumn{1}{c}{10\%}&    -3.13 &    -2.99 &    -2.81 &    -2.47 &    -2.04 \\
\multicolumn{1}{c}{20\%}&    -3.05 &    -2.91 &    -2.75 &    -2.46 &    -2.19 \\
\multicolumn{1}{c}{30\%}&    -2.90 &    -2.76 &    -2.62 &    -2.41 &    -2.28 \\
\multicolumn{1}{c}{40\%}&    -2.70 &    -2.59 &    -2.49 &    -2.36 &    -2.32 \\
\multicolumn{1}{c}{50\%}&    -2.52 &    -2.44 &    -2.39 &    -2.31 &    -2.33 \\
        \bottomrule
    \end{tabularx}
    \label{tab:tab_gini_tot_sr15}
\end{table}
\FloatBarrier

\justifying
\section{Mechanisms}
\label{sec:mechanisms}
\noindent Why does homeownership improve life-cycle wealth and welfare? We argue that homeownership mitigates two risks households face over the life cycle: investment risk and housing-cost risk. Together, these channels drive the consumption-equivalent premium of homeownership. This section further shows that the magnitude and timing of the gains depend on two additional features of the household problem: how well housing aligns resources with consumption needs across the life cycle, and how mortgage leverage trades off  early consumption for  later wealth. For ease of exposition, we focus on the baseline 15\% saving rate and report results for the full population of simulated households across the strategy grid.

\subsection{Risk Aversion}
\label{subsec:risk_aversion}

\begin{figure}[htbp]
    \centering
    \includegraphics[width=0.5\linewidth]{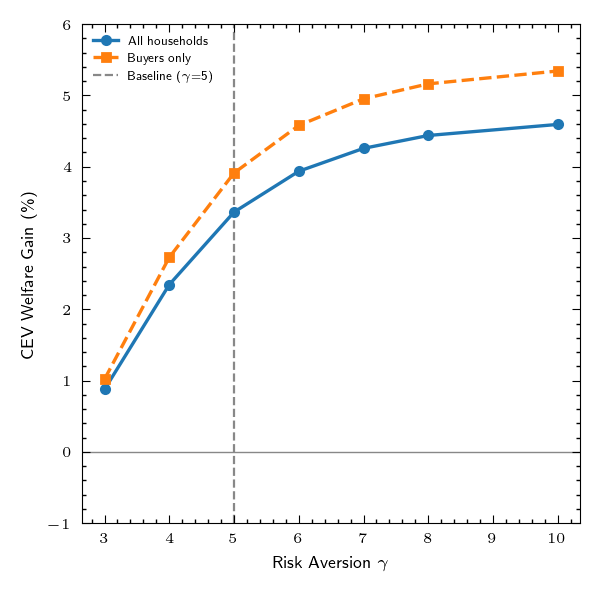}
    \caption{\textbf{Consumption-Equivalent Gains Across Risk-Aversion Coefficients}}
    \justifying{\footnotesize\noindent This figure reports average lifetime consumption-equivalent gains from homeownership as the risk-aversion coefficient $\delta$ varies from 3 to 10  for the full household population and for the subset of households that actually purchase a home. The baseline calibration is $\delta = 5$.}
    \label{fig:risk_aversion_sensitivity}
\end{figure}

\noindent A more risk-averse household assigns a higher utility cost to portfolio volatility and rent uncertainty and should value the insurance provided by homeownership more highly. Figure~\ref{fig:risk_aversion_sensitivity} confirms this prediction. As the risk-aversion coefficient $\delta$ rises from 3 to 10, the average consumption-equivalent gain from homeownership at the baseline saving rate increases monotonically at a decreasing rate---from roughly 1.0\% at $\delta = 3$ to 3.4\% at $\delta = 5$ and 4.6\% at $\delta = 10$, with the steepest increase concentrated between $\delta = 3$ and $\delta = 6$.


Since our main results compare strategies rather than realized outcomes, a household adopting a homeowner strategy may not end up purchasing a home. We therefore also report consumption-equivalent gains conditional on home purchase but stress that this conditions on an endogenous outcome: Households that buy are disproportionately those with favorable income and return histories, so the conditional gain reflects selection as well as ownership. At the baseline $\delta = 5$, the buyers-only gain reaches approximately 3.9\% compared with 3.4\% for the full population, and the gap widens at higher levels of risk aversion. Overall, our findings confirm that the welfare advantage of homeownership access increases with risk aversion, consistent with risk aversion as a key determinant of homeownership's value in the  household utility function.

The welfare gains documented above reflect two  insurance channels embedded in the homeownership contract. On the asset side, homeownership mitigates investment risk: Housing substitutes for part of the equity portfolio with an asset that is less volatile and less correlated with equity returns, reducing portfolio volatility and trimming the left tail of outcomes. On the liability side, it mitigates housing-cost risk: Ownership converts a stochastic and persistent rental expense into a stable owner housing cost. The next two subsections develop each channel in turn.

\justifying
\subsubsection{Housing as a Better Bond-Like Asset}
\label{subsubsec:housing_as_a_better_bond_like_asset}
\begin{figure}[!ht]
    \centering
    \includegraphics[width=\linewidth]{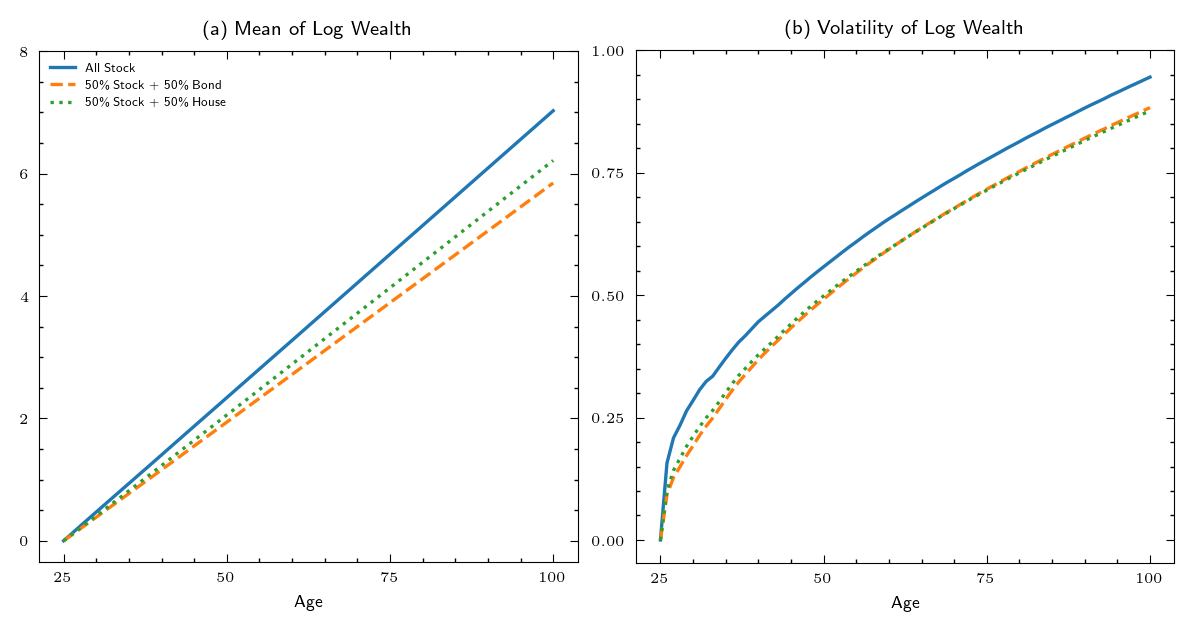}
    \caption{\textbf{Mean and Standard Deviation of Log Wealth over Simplified Life Cycle}}
    \justifying{\footnotesize\noindent This figure presents age profiles of log wealth from 1~million stationary block-bootstrap simulations, abstracting from labor-income, retirement, and other life-cycle factors. Initial wealth is normalized to one at age 25, and households are followed to age 100. We compare three investment strategies: all-equity, a blend of 50\% stocks and 50\% bonds, and a blend of 50\% stocks and 50\% housing. The left panel plots mean log wealth by age, and the right panel plots the cross-sectional volatility of log wealth.}
    \label{fig:portfolio_log_wealth_strategies}
\end{figure}

\noindent We begin with the asset side. For risk-averse households, assets that reduce consumption volatility while preserving long-run returns are particularly valuable. We show that housing occupies a distinct position between the two standard asset classes: It delivers substantially higher long-run returns than government bonds at comparable volatility  while exhibiting far less volatility than
equities. Figure~\ref{fig:portfolio_log_wealth_strategies} compares housing's long-run performance with government bonds and equities. We simulate 1~million households with initial wealth normalized to one, abstracting from labor income, retirement, and other life-cycle factors, and follow them to age 100. This simplified simulation is distinct  from the analysis in Section~\ref{sec:main_results} and isolates asset characteristics from life-cycle considerations.

The results show that replacing bonds with housing in a balanced portfolio substantially narrows the wealth gap with an all-equity strategy. At age 100, mean log wealth reaches approximately 6.3 for the blend of 50\% stocks and 50\% housing, compared with 5.8 for the blend of 50\% stocks and 50\% bonds   and 7.1 for the all-equity strategy. The housing-equity blend thus closes roughly 40\% of the gap between the bond-equity blend and all-equity, indicating that housing offers substantially better long-run returns than bonds while retaining a comparable diversification role in a balanced portfolio.
The volatility comparison reinforces this conclusion. At age 100, the housing-equity blend generates a log-wealth standard deviation nearly identical to that of the bond-equity blend (approximately 0.87 versus 0.90) and meaningfully below the all-equity strategy (0.94). Housing therefore dominates bonds on both dimensions of the risk-return trade-off: Substituting housing for bonds in a balanced portfolio delivers roughly 0.5 additional log-wealth units without any increase in long-run volatility.


In short, as an asset class, housing dominates bonds in the long-run risk-return trade-off, delivering substantially higher returns at comparable volatility, consistent with time-series and mean-variance evidence in the literature \citep{goetzmann1993single,jorda2019rate}. The asset-side channel, however, is only one side of the insurance mechanism. Homeownership also transforms the household's liability profile, converting an uncertain stream of future rents into a stable ownership cost, the channel we examine in the next subsection.


\subsubsection{Ownership Costs and Rent Risk}
\label{subsubsec:cost_stability}
\noindent We turn now to the liability side. In our simulation, the rent-to-price ratio fluctuates substantially, exposing renters to volatile rental costs. Because rent shocks are also highly persistent, they accumulate into large swings in lifetime rental expenditure.\footnote{\citet{sinai2005owner} document a serial correlation of approximately 0.85 in U.S. rents using data at the metropolitan-statistical-area level.} Owners, by contrast, face a far more predictable cost stream. Maintenance costs are 2\% of home value, and annual mortgage payments are predetermined under the fixed-rate contract \citep{cocco2020aging,yao2004optimal,nakajima2017reverse}. Homeownership thus converts a variable liability into a fixed one, and the welfare gain reflects  insurance value embedded in the liability side of the balance sheet.

\begin{figure}[!t]
    \centering
    \includegraphics[width=0.5\linewidth]{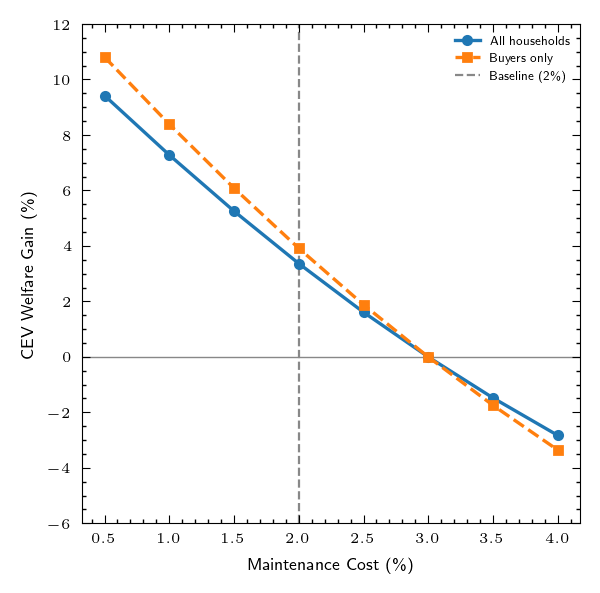}
    \caption{\textbf{Consumption-Equivalent Gains Across Maintenance-Cost Assumptions}}
    \justifying{\footnotesize\noindent This figure reports lifetime consumption-equivalent gains as the annual maintenance cost varies from 0.5\% to 4\% of home value for the full household population and for the subset of households that actually purchase a home. The baseline calibration is 2\%. }
    \label{fig:maintenance_cost_sensitivity}
\end{figure}

Figure~\ref{fig:maintenance_cost_sensitivity} shows that this insurance value remains positive across the empirically relevant range of maintenance costs. At the 2\% baseline, homeownership delivers an average consumption-equivalent gain of approximately 3.4\% across the purchasing-strategy grid and 3.9\% among buyers, and the gain declines approximately linearly as the maintenance rate rises. The zero crossing occurs only at approximately 3\% annual maintenance, meaning homeownership loses its consumption-equivalent advantage only when the cost of ownership stands 50\% above the baseline calibration. This is an extreme scenario well outside empirical estimates of typical maintenance expenditure \citep{cocco2020aging,yao2004optimal,nakajima2017reverse}. The robustness of the welfare gain across cost calibrations supports the interpretation that the consumption-equivalent premium arises from risk reduction rather than from a knife-edge cost advantage.

\justifying
\subsection{Intertemporal Substitution}
\label{subsec:intertemporal_substitution}
\noindent The average welfare gains documented above conceal a striking intertemporal pattern: Homeowners accumulate substantially less wealth than renters throughout working life yet incur almost no welfare loss in doing so, and the gains from ownership arrive disproportionately late in the life cycle. Figure~\ref{fig:age_profile_best_welfare} traces this profile by age for the welfare-maximizing strategies at each saving rate.\footnote{The CEV is calculated by age following Equation~\ref{eq: equiw}.} The columns vary the saving rate, and the rows report net worth (top) and CEV (bottom).

At the baseline 15\% saving rate, consumption-equivalent gains build gradually from near zero in early working life, peak at approximately 5.6\% at age 65, and remain in the 3.3\% to 3.9\% range thereafter. The consumption-equivalent gain from homeownership is therefore not a one-time transfer but a permanent upward shift in household utility that lasts into old age. Raising the saving rate to 20\% preserves this shape and amplifies it, with a peak of about 6.6\%. The mechanism is precautionary: A larger financial buffer sustains mortgage service and reduces the risk of forced liquidation during downturns.

The net-worth profiles reveal the intertemporal cost of these gains. Owners hold less net worth than renters during working life, and the shortfall is modest at a 15\% saving rate but sharper at 20\%. Owners overtake renters only in retirement, reaching approximately \$4.8 million against the renter's \$4.4 million by age 90. Homeownership thus reallocates resources across the life cycle, trading lower consumption and wealth during working years for higher, more stable welfare in retirement. This intertemporal exchange is thus another central determinant of the life-cycle value of homeownership.

\justifying
\subsection{Leverage-Induced Substitution Between Consumption and Timing of Purchase}
\label{subsec:leverage_consumption_timing_substitution}
\noindent The preceding subsections established that the welfare gain from homeownership reflects intertemporal substitution across the life cycle together with risk reduction on both sides of the household balance sheet. A separate question is how the household should optimally structure the purchase itself: how much leverage to take and when to enter the market. We show that these two margins interact through a leverage-induced trade-off between early  consumption and early ownership.

\begin{figure}[htbp]
    \centering
    \includegraphics[width=0.5\linewidth]{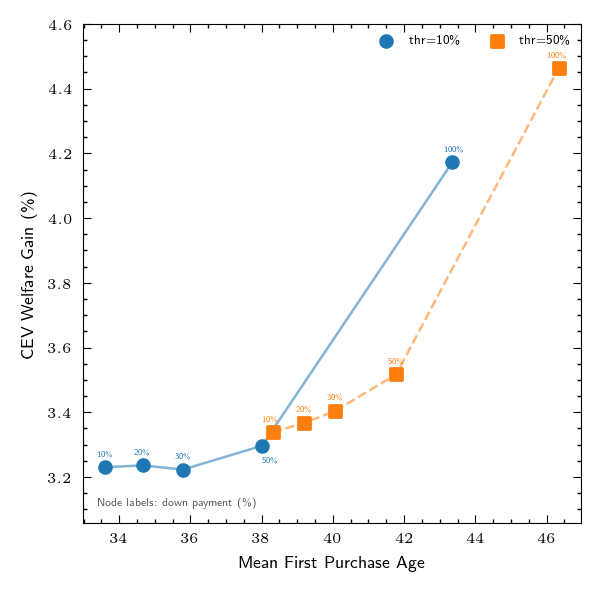}
    \caption{\textbf{Consumption-Equivalent Gains Versus Mean Home-Purchase Age}}
    \justifying{\footnotesize\noindent This figure reports lifetime consumption-equivalent gains against the mean age of first home purchase for the low-threshold 10\% series and the high-threshold 50\% series, with node labels identifying the down payment share. Each point represents one homeowner strategy.}
    \label{fig:cev_vs_purchase_age}
\end{figure}

Figure~\ref{fig:cev_vs_purchase_age} plots consumption-equivalent gains against mean purchase age at the 15\% saving rate. The two series correspond to the low (10\%) and high (50\%) purchase thresholds, and node labels identify the down payment share. Cash-purchase strategies appear at the far right of the figure and deliver the highest consumption-equivalent gains, while more leveraged strategies cluster at younger purchase ages with lower but still positive gains. 

Three patterns stand out. First, within the leveraged region, gains rise with purchase age: Each additional year of purchase delay is associated with roughly 0.1\% to 0.2\% additional consumption-equivalent gain, so  leveraged strategies deliver gains between 3.2\% and 3.5\% depending on the down payment share and threshold combination. Second, the choice of purchase threshold shifts the entire schedule outward, moving mean purchase ages from the mid-thirties under the low threshold into the early forties under the high threshold. Third, and most striking, there is a discrete jump at the transition from leveraged to cash purchase: Moving from the highest-leverage strategy to debt-free purchase raises the consumption-equivalent gain by roughly 0.9\% under the low threshold and 1.0\% under the high threshold, a magnitude 5 to 10 times larger than the gain from an equivalent year of delay within the leveraged region.

The mechanism behind the first two patterns is intuitive: Households that delay accumulate larger financial buffers, carry a lower mortgage burden per dollar of accumulated equity, and enjoy smoother retirement consumption. In contrast, the third pattern reflects a distinct mechanism, namely the outright elimination of mortgage-induced consumption suppression. The roughly 1\% welfare gap between leveraged and cash strategies therefore represents the consumption-equivalent cost that households accept to own their home approximately 10 years earlier.

Our model captures welfare through a purely monetary lens, so nonmonetary benefits of homeownership, such as stability, community ties, and school quality, are absent. Were they included, the welfare case for leveraged and earlier purchase would strengthen substantially, and the true value of accelerated homeownership would exceed what the monetary comparison alone reveals.

\justifying
\subsection{Liquidity}
\label{subsec:liquidity}
\noindent The welfare gains documented above come at a cost in liquidity during working life. Down payments and mortgage service divert resources from the stock portfolio, so homeowners enter retirement holding smaller liquid financial portfolios than their saving-rate-matched renter counterparts. Table~\ref{tab:tab_fin_wealth_tot_sr15} quantifies this effect at the baseline 15\% saving rate. Panel~A shows that homeowners retire with 26.1\% to 38.4\% less liquid financial wealth than their saving-rate-matched renter counterparts across the strategy grid, with the shortfall widening as the down payment share rises. This crowding-out pattern is consistent with the life-cycle housing literature \citep{becker2010outstanding, cocco2005housing, vestman2019limited, yao2004optimal}.

\begin{table}[!t]
   \justifying{\noindent \footnotesize This table reports liquidity outcomes from stationary block-bootstrap simulations of 1~million household life cycles at the baseline 15\% saving rate. The homeowner and the matched renter face identical simulated labor-income path, asset-return, inflation, and survival paths, and both follow the same saving rule after fixed costs. The renter rents for life and invests only in stocks. The homeowner purchases once financial wealth covers the ``Down Payment'' plus an additional purchase ``Threshold,'' both as percentages of home value. Financial wealth comprises the liquid financial portfolio and excludes housing equity. Each entry reports the percentage difference in liquid financial wealth between the homeowner and the matched all-equity renter benchmark; negative values indicate that homeowners hold less liquid financial wealth than the matched all-equity renters. Panel~A reports the difference at retirement (age 65), and Panel~B reports the difference at death. All values are in percent.\vspace{4pt}}\par
    \centering
    \caption{\textbf{Financial Wealth at Retirement and Death}}
    \footnotesize
    \begin{tabularx}{\textwidth}{p{.15\linewidth}YYYYY}
        \toprule
\multicolumn{6}{c}{Panel A: Financial Wealth Change at Retirement} \\
\midrule
& \multicolumn{5}{c}{Down Payment} \\
\cmidrule{2-6}
\multicolumn{1}{c}{Threshold}
& 10\% & 20\% & 30\% & 50\% & 100\% \\
\midrule
\multicolumn{1}{c}{10\%}&   -27.44 &   -29.73 &   -31.67 &   -34.62 &   -38.43 \\
\multicolumn{1}{c}{20\%}&   -27.80 &   -29.72 &   -31.30 &   -33.69 &   -36.88 \\
\multicolumn{1}{c}{30\%}&   -27.77 &   -29.30 &   -30.57 &   -32.48 &   -35.37 \\
\multicolumn{1}{c}{40\%}&   -27.16 &   -28.39 &   -29.43 &   -31.07 &   -33.90 \\
\multicolumn{1}{c}{50\%}&   -26.13 &   -27.16 &   -28.06 &   -29.58 &   -32.49 \\
\midrule
\multicolumn{6}{c}{Panel B: Financial Wealth Change at Death} \\
\midrule
& \multicolumn{5}{c}{Down Payment} \\
\cmidrule{2-6}
\multicolumn{1}{c}{Threshold}
& 10\% & 20\% & 30\% & 50\% & 100\% \\
\midrule
\multicolumn{1}{c}{10\%}&    -9.29 &   -10.97 &   -12.41 &   -14.59 &   -17.08 \\
\multicolumn{1}{c}{20\%}&    -9.85 &   -11.26 &   -12.44 &   -14.17 &   -16.21 \\
\multicolumn{1}{c}{30\%}&   -10.17 &   -11.30 &   -12.23 &   -13.62 &   -15.40 \\
\multicolumn{1}{c}{40\%}&   -10.10 &   -11.00 &   -11.76 &   -12.97 &   -14.65 \\
\multicolumn{1}{c}{50\%}&    -9.73 &   -10.49 &   -11.17 &   -12.28 &   -13.96 \\
        \bottomrule
    \end{tabularx}
    \label{tab:tab_fin_wealth_tot_sr15}
\end{table}

The liquidity gap narrows after retirement as homeowners gradually unlock home equity through sale or reverse mortgage. Panel~B shows that by death the financial wealth gap has shrunk to 9.3\% to 17.1\%, roughly one-third to one-half of its size at retirement. Because overall life-cycle welfare is stronger for homeowners, as documented in Section~\ref{sec:main_results}, the early liquidity loss represents the price households pay for lower portfolio risk and a stronger late-life balance sheet. The illiquid home equity accumulated during the working periods also converts into bequest value at termination, as Table~\ref{tab:tab_dissect_welfare_tot_sr15} shows.

The liquidity channel therefore complements the mechanisms discussed above: Risk aversion, intertemporal smoothing, and leverage together determine why homeownership improves welfare, while liquidity determines the working-age cost households bear to obtain that improvement.



\section{Second-Home Ownership}
\label{sec:second_home_ownership}
\noindent Households that have paid off their first-home mortgage, hold no existing second home, and meet joint income and wealth criteria are eligible to acquire a second home in the simulation.\footnote{The income criterion requires that current labor income net of minimum consumption and total maintenance costs for both homes covers the second-home mortgage payment. The wealth criterion requires financial assets to cover the outstanding first-home mortgage balance plus the second-home down payment, purchase threshold, and transaction costs.} Because the second-home option layers onto every first-home strategy rather than onto any single baseline calibration, we measure its marginal effect by averaging outcomes across the 25 single-home strategies before and after the option is introduced, at each second-home threshold. Table~\ref{tab:tab_h2_sr15} reports wealth at death, CEV gain, financial-ruin reduction, and Gini reduction, all relative to the saving-rate-matched all-equity renter benchmark.\footnote{Our calculation of the economic value on the second home excludes the depreciation that can be applied to a typical investment house, so the reported gains understate the value of a second home held as a rental.}


\begin{table}[!t]
    \justifying{\noindent \footnotesize This table reports second-home outcomes from stationary block-bootstrap simulations of 1~million household life cycles at the baseline 15\% saving rate. The underlying single-home strategies span the full 25-cell threshold-by-down-payment grid. The first row reports the average single-house outcome across that grid. Each remaining row reports the average outcome after imposing the indicated second-house additional purchase threshold (``Second-Home Threshold'') on those same underlying strategies. The second-home threshold is measured as a percentage of home value. ``Wealth at Death'' reports the mean dollar value wealth at death for the homeowner strategies. Consumption-equivalent variation (CEV) is the percentage difference between the homeowner and renter benchmark strategies in the constant consumption levels that replicate each strategy's realized lifetime utility, averaged across households. A positive CEV is a consumption-equivalent gain. Values in ``Consumption-Equivalent Gain'' are in percent. ``Financial-Ruin Reduction'' and ``Gini Reduction'' report level differences in percent. More negative values indicate larger reductions in financial ruin and wealth inequality. Thus, an entry of $-12$ for financial ruin means the homeowner ruin rate is 0.12 lower than the renter benchmark, and an entry of $-3$ for Gini means the homeowner Gini is 0.03 lower.\vspace{4pt}}\par
    \centering
    \caption{\textbf{Second-Home Outcomes}}
    \footnotesize
    \begin{tabularx}{\textwidth}{p{.20\linewidth}YYYY}
        \toprule
        Benchmark
        & Wealth at Death & Consumption-Equivalent Gain at Death & Financial-Ruin Reduction & Gini Reduction \\
        \midrule
        Single-Home Average & 4477865.60 & 3.57 & -11.50 & -2.57 \\
        \midrule
        Second-Home Threshold
        & Wealth at Death & Consumption-Equivalent Gain at Death & Financial-Ruin Reduction & Gini Reduction \\
        \midrule
        10\% & 4785937.30 & 1.15 & -12.16 & -3.59 \\
        20\% & 4782165.58 & 1.52 & -12.11 & -3.65 \\
        30\% & 4780568.23 & 1.88 & -12.06 & -3.65 \\
        40\% & 4774881.64 & 2.22 & -12.06 & -3.62 \\
        50\% & 4781205.98 & 2.55 & -11.97 & -3.60 \\
        \bottomrule
    \end{tabularx}
    \label{tab:tab_h2_sr15}
\end{table}

\justifying
Adding a second home raises terminal wealth and strengthens downside protection across the strategy grid. Mean wealth at death rises from \$4.48 million under the single-house average to roughly \$4.79 million at a second-house threshold of 10\%, a gain of about 6.9\%. Across the remaining thresholds, the level stays within a narrow \$4.77 million to \$4.79 million range. This weak sensitivity to the threshold indicates that once a household is financially strong enough to acquire a second property, the wealth gain holds under a broad range of entry rules rather than only under one favorable calibration. The financial-ruin risk relative to the matched renter benchmark also falls further, from an 11.50\% reduction under the single-house average to between 11.97\% and 12.16\% under second-home ownership. The additional property therefore strengthens the late-life balance sheet and provides an extra buffer against financially distressed states.


Consumption-equivalent gains, by contrast, decline once a second home is added. At every second-house threshold, they remain positive but fall well below the 3.57\% single-house average, ranging from 1.15\% at the 10\% threshold to 2.55\% at the 50\% threshold. The monotone increase with the threshold reflects the liquidity channel: A higher threshold delays entry into the second property and preserves more liquid savings, mitigating the consumption compression caused by an additional illiquid real estate position. Even the best-performing second-house strategy falls short of the single-house average  because the extra property raises terminal wealth while diverting resources from the liquid portfolio that supports consumption smoothing and retirement spending.


The distributional effect, however, strengthens. The Gini reduction at retirement rises from 2.57\% under the single-house average to between 3.59\% and 3.65\% under second-house strategies, with the strongest compression around the 20\% to 30\% threshold. The second property therefore deepens the inequality-compression effect of housing even as it lowers average life-cycle consumption-equivalent gains. 

Taken together, the second-home results point to a clear portfolio trade-off. Additional housing exposure improves late-life wealth accumulation, lowers financial-ruin risk, and compresses retirement wealth inequality, but the added illiquidity prevents these balance-sheet gains from fully translating into higher average consumption-equivalent gains over the life cycle.

\section{Conclusion}
\label{sec:conclusion}
\noindent Contrary to the expert advice to rent and invest the discretionary savings in financial assets, homeownership remains a valuable investment when the alternatives are the common benchmark investment strategies. Using stationary block-bootstrap simulations drawn from 150 years of data across 16 countries, we show that homeowner strategies generally enhance life-cycle wealth and welfare. At the baseline saving rate of 15\%, homeowners realize consumption-equivalent gains up to 4.5\% relative to the most stringent benchmark, the all-equity strategy. Gains are even larger against the common investment benchmarks, such as the balanced 60/40 allocation, 100-minus-age, and TDF strategies. 

The wealth advantage peaks under low down payments, whereas the consumption-equivalent advantage is greatest for  cash purchases. Consumption-equivalent gains also rise with the saving rate, as stronger financial buffers sustain mortgage service and support retirement consumption. Homeownership further preserves wealth, lowers downside risks, and reduces inequality. These gains reflect four mechanisms. First, housing substitutes for bonds at a higher return and converts volatile rent into a fixed owner housing cost. Second, homeownership shifts consumption toward retirement. Third, mortgage leverage trades current consumption for earlier ownership. Buying 10 years sooner at a 10\% down payment costs approximately 1\% in lifetime consumption-equivalent terms. Fourth, the housing position exchanges early liquidity for late-life balance-sheet strength. 

Within groups following the same life-cycle investment strategy, the economic benefits of homeownership are   shaped by heterogeneous household profiles and the timing of home purchases. Extending the framework to a second property reveals a clear portfolio trade-off: Additional housing exposure strengthens the late-life balance sheet and heightens inequality compression, but the illiquidity it entails limits the pass-through to lifetime consumption-equivalent gains.



\clearpage


\printbibliography 

\clearpage

\bookmark[level=1,dest=\hyperget{anchor}{fig:homeownership_strategy_surface}]{Figures}
\bookmark[level=2,dest=\hyperget{anchor}{fig:homeownership_strategy_surface}]{Optimal Life-Cycle Homeowner Strategy}
\bookmark[level=2,dest=\hyperget{anchor}{fig:age_profile_best_welfare}]{Household Age Profile over the Life Cycle}
\bookmark[level=2,dest=\hyperget{anchor}{fig:labor_income_fan_chart}]{Labor Income by Age}
\bookmark[level=2,dest=\hyperget{anchor}{fig:risk_aversion_sensitivity}]{Consumption-Equivalent Gains Across Risk-Aversion Coefficients}
\bookmark[level=2,dest=\hyperget{anchor}{fig:portfolio_log_wealth_strategies}]{Mean and Standard Deviation of Log Wealth over Simplified Life Cycle}
\bookmark[level=2,dest=\hyperget{anchor}{fig:maintenance_cost_sensitivity}]{Consumption-Equivalent Gains Across Maintenance-Cost Assumptions}
\bookmark[level=2,dest=\hyperget{anchor}{fig:cev_vs_purchase_age}]{Consumption-Equivalent Gains Versus Mean Home-Purchase Age}
\bookmark[level=1,dest=\hyperget{anchor}{tab:return_summary_statistics}]{Tables}
\bookmark[level=2,dest=\hyperget{anchor}{tab:return_summary_statistics}]{Summary Statistics}
\bookmark[level=2,dest=\hyperget{anchor}{tab:param_household}]{Parameterization}
\bookmark[level=2,dest=\hyperget{anchor}{tab:tab_life_cycle_gains_tot}]{Life-Cycle Gains}
\bookmark[level=2,dest=\hyperget{anchor}{tab:tab_benchmark_all_strategies_tot_reorg}]{Life-Cycle Gains Versus Common Renter Benchmarks}
\bookmark[level=2,dest=\hyperget{anchor}{tab:tab_dissect_welfare_tot_sr15}]{Welfare Decomposition}
\bookmark[level=2,dest=\hyperget{anchor}{tab:tab_hetero_avg_sr15}]{Heterogeneity}
\bookmark[level=2,dest=\hyperget{anchor}{tab:tab_life_cycle_gains_regional_sr15}]{Regional Data Evidence}
\bookmark[level=2,dest=\hyperget{anchor}{tab:tab_wealth_default_tot_sr15}]{Downside Risk Outcomes}
\bookmark[level=2,dest=\hyperget{anchor}{tab:tab_gini_tot_sr15}]{Gini Change at Retirement}
\bookmark[level=2,dest=\hyperget{anchor}{tab:tab_fin_wealth_tot_sr15}]{Financial Wealth at Retirement and Death}
\bookmark[level=2,dest=\hyperget{anchor}{tab:tab_h2_sr15}]{Second-Home Outcomes}


\end{document}